\documentclass[9pt]{elife}

\usepackage[version=4]{mhchem}
\usepackage{siunitx}
\usepackage{amsthm}

\usepackage[percent]{overpic}
\usepackage{textcomp}

\DeclareSIUnit\Molar{M}

\title{From Single Neurons to Behavior in the Jellyfish \textit{Aurelia aurita}}

\author[1*]{Fabian Pallasdies}
\author[1]{Sven Goedeke}
\author[1]{Wilhelm Braun}
\author[1*]{Raoul-Martin Memmesheimer}
\affil[1]{Neural Network Dynamics and Computation, Institute of Genetics, University of Bonn, Bonn, Germany}

\corr{fabianpallasdies@gmail.com}{FP}
\corr{rm.memmesheimer@uni-bonn.de}{RMM}

\begin{document}

\maketitle

\begin{abstract}

Jellyfish nerve nets provide insight into the origins of nervous systems, as both their taxonomic position and their evolutionary age imply that jellyfish resemble some of the earliest neuron-bearing, actively-swimming animals.
Here we develop the first neuronal network model for the nerve nets of jellyfish. Specifically,
we focus on the moon jelly \textit{Aurelia aurita} and the control of its energy-efficient swimming motion.
The proposed single neuron model disentangles the contributions of different currents to a spike. The network model identifies factors ensuring non-pathological activity and suggests an optimization for the transmission of signals. After modeling
the jellyfish's muscle system and its bell in a hydrodynamic environment, we explore the swimming elicited by neural activity. We find that different delays between nerve net activations lead to well-controlled, differently directed movements. Our model bridges the scales from single neurons to behavior, allowing for a comprehensive understanding of jellyfish neural control.
\end{abstract}

\section{Introduction}
\subsection{Modeling Jellyfish}

Understanding how neural activity leads to behavior in animals is a central goal in neuroscience. 
Since jellyfish are anatomically relatively simple animals with a limited behavioral repertoire \citep{Albert_2011}, modeling their nervous system opens up the possibility to achieve this goal.

Cnidarians (in particular jellyfish) and ctenophores (comb jellies) are the only non-bilaterian animal phyla with neurons. While their phylogenetic position is still not entirely resolved, evidence suggests that cnidarians are our most distant relatives with homologous neurons and muscles \citep{Steinmetz_Kraus_Larroux_Hammel_Amon-Hassenzahl_Houliston_Woerheide_Nickel_Degnan_Technau_2012, Marlow_Arendt_2014, Moroz_Kohn_2016}. Well-preserved fossils of medusozoa from the Cambrian \citep{Cartwright_Halgedahl_Hendricks_Jarrard_Marques_Collins_Lieberman_2007} and evidence for medusoid forms from the Ediacaran \citep{Van_Iten_de_Moraes_Leme_Simoes_Marques_Collins_2006} indicate that jellyfish are evolutionary old. These findings and their anatomical simplicity suggest that they
are similar to the earliest neuron-bearing, actively swimming animals.
Their study should therefore yield insight into the earliest nervous systems and behaviors.

The present study focuses on the neuro-muscular control of the swimming motion in a true (scyphozoan) jellyfish in the medusa stage of development.
Specifically, incorporating available experimental observations and measurements, we develop a bottom-up multi-scale computational model of the nerve nets, the muscle system and the bell of the moon jelly \textit{Aurelia aurita}. Using fluid-structure hydrodynamics simulations, we then explore how the nervous system generates and shapes different swimming motions.

Before presenting our results, we review the current knowledge on the nervous systems of jellyfish in the following introductory sections, also highlighting specific open questions that further motivate our study.

\subsection{Nervous Systems of Scyphomedusae}

The nervous system of scyphozoan jellyfish consists of several neuronal networks, which are distributed over the entire jellyfish bell, the tentacles and the endoderm \citep{Schafer_1878,Passano_Passano_1971}. The only obvious points of concentration of a larger number of neurons are the rhopalia, small sensory structures  of which there are usually eight distributed around the margin of the bell \citep{Nakanishi_Hartenstein_Jacobs_2009}.

Much of the current knowledge on the inner workings of these nerve nets, in particular concerning the control of the swimming muscles, was already formulated by George Romanes in the 19th century \citep{Romanes_1885}. During a swimming motion almost all the subumbrellar muscles
contract synchronously and push the jellyfish forward. In a series of cutting experiments, Romanes destroyed and removed parts of the umbrella. He found that the contraction usually starts at one of the rhopalia and propagates around almost arbitrary cuts in the subumbrella. Furthermore, Romanes observed two different types of contraction waves:
a fast, strong wave leading to the regular swimming motion and a slower wave, at about half the speed, which was so weak that one could hardly see it activate the swimming muscles. When a slow contraction wave originating somewhere on the outer margin of the jellyfish umbrella reached a rhopalium, a fast excitation wave emerged from that rhopalium after a short delay.

With advancing neurobiological methods, Romanes' observations
were later verified and expanded \citep{Passano_1965, Satterlie_2002}. This led to the identification of two different nerve nets, the motor nerve net (MNN) and the diffuse nerve net (DNN), being responsible for the fast and slow contraction wave, respectively.

\subsection{The Motor Nerve Net}
The motor nerve net extends over the subumbrella and consists of large neurons with usually two neurites \citep{Schafer_1878, Anderson_Schwab_1981,Satterlie_2002}. The neurons function in basically the same manner as neurons with chemical synapses in higher animals \citep{Anderson_Schwab_1983, Anderson_1985}.

The MNN is throughconducting in the sense that if a small number of neurons is activated,
a wave of activation spreads over the entire network, leading to a series of neuronal discharges. The conduction speed is between $45~\text{cm/s}$ and $1~\text{m/s}$ \citep{Horridge_1956, Passano_1965}. The activation is preserved even if large parts of the network are destroyed and
generates Romanes' fast contracting wave in the swimming muscles \citep{Horridge_1956}. 

Spontaneous waves in the MNN are initiated by pacemakers in each of the rhopalia \citep{Passano_1965}.
After firing, the wave initiating pacemaker resets and the other ones reset due to the arriving MNN activity. \cite{Horridge_1959} showed that sensory input modulates the pacemaker activity. This may be one of the main mechanisms of sensory integration and creation of controlled motor output in the jellyfish.

In studies that investigated the electrophysiology of the MNN in detail, remarkable features have been observed.
First, even though the synapses seem to be exclusively chemical they are symmetrical, both morphologically and functionally. Both sides of the synaptic cleft have a similar structure containing vesicles as well as receptors \citep{Horridge_Mackay_1962}. Electrical synapses have not been found, neither through staining nor in electrophysiological experiments \citep{Anderson_Schwab_1981, Anderson_1985,Anderson_Spencer_1989}. In particular, the neurites do not differentiate into axon and dendrite. \cite{Anderson_1985} directly showed that the conduction is bidirectional.
Second, synapses are strong, usually creating an excitatory postsynaptic potential (EPSC) that induces an  action potential (AP) in the receiving neuron \citep{Anderson_1985}.
This fits with the observed robust throughconductance of the MNN found during \textit{in vivo} cutting experiments \citep{Horridge_1954b}.
However, it also raises the question why symmetrical synapses of such strength do not lead to repetitive firing in (sub-)networks of neurons or even to epileptic dynamics.

\subsection{The Diffuse Nerve Net}
Historically, any neuron not associated with the MNN or the rhopalia was categorized into the DNN, including the neurons in the manubrium and the tentacles \citep{Horridge_1956}. We adopt the nomenclature of more recent studies, where the term DNN refers mostly to the throughconducting nerve net of the ex- and subumbrella, which does not directly interact with the MNN \citep{Arai_1997}. Little is known about the DNN's small neurons and its synapses.
The conduction speed of activity waves ($15~\text{cm/s}$) along the subumbrella is smaller than in the MNN \citep{Passano_1973}.

\cite{Horridge_1956} was the first to suggest that innervation of the swimming muscles via the DNN with its slower time scale may allow for a different activation pattern and thereby induce a turning motion. This could be achieved by a simultaneous versus a successive arrival of MNN- and DNN-generated contraction waves on two sides of the animal. In \textit{Aurelia}, however, no visible contraction of the regular swimming muscles after DNN excitation was observed \citep{Horridge_1956}.
Still, the DNN might influence them by amplifying the impact of the MNN activity
as it was measured in other jellyfish \citep{Passano_1965, Passano_1973}. In addition there is a small band of radial muscles on the marginal angles of \textit{Aurelia}, which contracts only during a turning motion on the inside of the turn \citep{Gemmell_Troolin_Costello_Colin_Satterlie_2015}. These muscles appear to be innervated by the DNN and influence the swimming motion significantly \citep{Flammang_Lauder_Troolin_Strand_2011}.

In accordance with the idea of a coupled activation of DNN and MNN,
DNN activity can activate the MNN indirectly via a rhopalium. The delay observed between DNN activity arrival and the initiation of the MNN activation
are highly variable \citep{Passano_1965, Passano_1973}.
Apart from this, the DNN does not directly interact with the MNN \citep{Horridge_1956}. 
Some behavioral \citep{Horridge_1956, Gemmell_Troolin_Costello_Colin_Satterlie_2015} and anatomical \citep{Nakanishi_Hartenstein_Jacobs_2009} evidence suggests that a rhopalium might activate the DNN together with the MNN in response to a strong sensory stimulus. 
These points indicate that each rhopalium is responsible for steering the animal by stimulating either one or both of the nerve nets. If and how the jellyfish can control its swimming motion beyond this is currently unknown.

\section{Results}
\subsection{A Model for Scyphozoan Neurons}
\label{sec:neuronmodel}
\subsubsection*{Model Construction and Comparison to Data}
We develop a biophysically plausible scyphozoan neuron model on the level of abstraction of Hodgkin-Huxley type single compartment models. These describe the actual voltage and current dynamics well and there is sufficiently detailed electrophysiological data available to fit such a model, obtained from \textit{Cyanea capillata} \citep{Anderson_1989}. Furthermore, dynamical mechanisms are not obscured by overly many variables and the models lend themselves to fast simulations of medium size neural networks, with several thousands of neurons. 

We incorporate the voltage-dependent transmembrane currents observed for scyphozoan MNN neurons by \cite{anderson1987properties, Anderson_1989} and fit the model parameters to the voltage-clamp data presented there (see Methods). 
The results of the fitting procedure are shown in Fig.~\ref{fig:Neuronmodel} A. The current traces of the biophysical model agree well with the measured ones, both qualitatively and quantitatively, for the broad experimentally explored range of clamping from $-20~\text{mV}$ to $+90~\text{mV}$ (step-size: $7.5~\text{mV}$, resting potential: $-70~\text{mV}$). The remaining unknown features of the model are the membrane capacitance and the synapse model. We choose them such that (i) the shape of excitatory postsynaptic potentials resembles the experimentally found one \citep{Anderson_1985},  (ii) the inflection point of an AP is close to $0~\text{mV}$ \citep{Anderson_Schwab_1983} and (iii) it takes approximately $2.5~\text{ms}$ for an AP to reach peak amplitude after stimulation via an EPSC (see Fig.~\ref{fig:Neuronmodel} B) \citep{Anderson_1989}.

\subsubsection*{Action Potentials and Synapses}
\begin{figure}[htb!]
    \begin{fullwidth}
    \centering
    \begin{overpic}[width=.9\paperwidth]{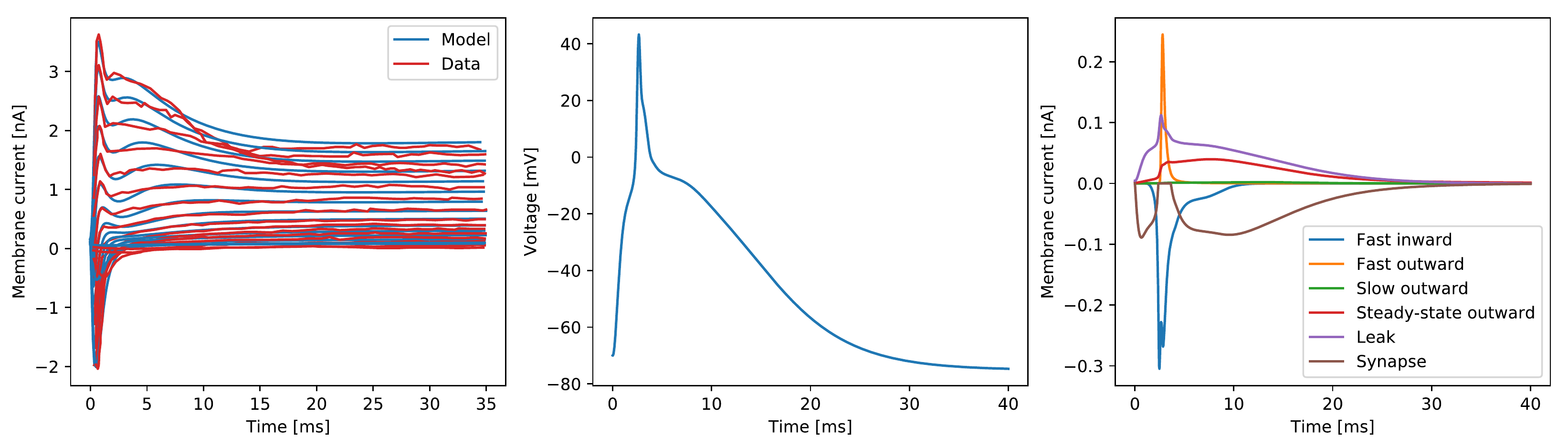}
        \put(0,26){A}
        \put(33,26){B}
        \put(67,26){C}
    \end{overpic}
    
    \caption[Neuron model]{\textbf{The biophysical model fitted to the voltage-clamp data.} 
    (\textit{A}) Comparison of our model dynamics with the voltage-clamp data \citep{Anderson_1989} that was used to fit its current parameters. The model follows the experimentally found traces.
    (\textit{B}) Membrane voltage of a neuron that is stimulated by a synaptic EPSC at time zero. The model neuron generates an action potential similar in shape to experimentally observed ones. (\textit{C}) The disentangled transmembrane currents during an action potential.

    }
    \label{fig:Neuronmodel}
    \end{fullwidth}
    
\end{figure}

Our model generates APs similar to the ones observed experimentally by \cite{Anderson_Schwab_1983}. It allows to quantitatively disentangle the contributions of the different transmembrane channel populations, see Fig.~\ref{fig:Neuronmodel}. Before an AP, the leak current dominates. After the voltage surpasses the inflection point, the fast transient in- and outward currents generate the voltage spike. During the spike the steady-state outward current activates and stays active during repolarization. The slow outward current does not activate, since it requires depolarizations beyond $+55~\text{mV}$ \citep{Anderson_1989}.

As experimentally observed in scyphozoan MNN neurons \citep{Anderson_1985}, our model EPSCs have fast initial rise, initially fast and subsequently slow decay and a single EPSC suffices to evoke an AP in a resting neuron \citep{Anderson_1985}. Furthermore, we incorporate the experimentally observed synaptic rectification: the synaptic current influx decays to zero when the voltage approaches the reversal potential ($+4~\text{mV}$) but does not reverse beyond (cf.~brown trace in Fig.~\ref{fig:Neuronmodel} C). Synaptic transmission is activated when a neuron reaches $+20~\text{mV}$ from below, which happens during spikes only. Since synapses in MNN neurons are symmetrical \citep{Anderson_1985, Anderson_Gruenert_1988}, we hypothesize that after transmitter release into the synaptic cleft, both pre- and postsynaptic neurons receive an EPSC. In our model, this ``synaptic reflux'' is responsible for a delayed repolarisation: the voltage stays near zero for several milliseconds after the fast return from the spike peak, see Fig.~\ref{fig:Neuronmodel}B. This is also visible in electrophysiological recordings \citep{Anderson_Schwab_1983, Anderson_1985}.

\subsubsection*{Refractory Period}
\begin{figure}[htb!]
   
     \centering
     \begin{overpic}[width=\textwidth]{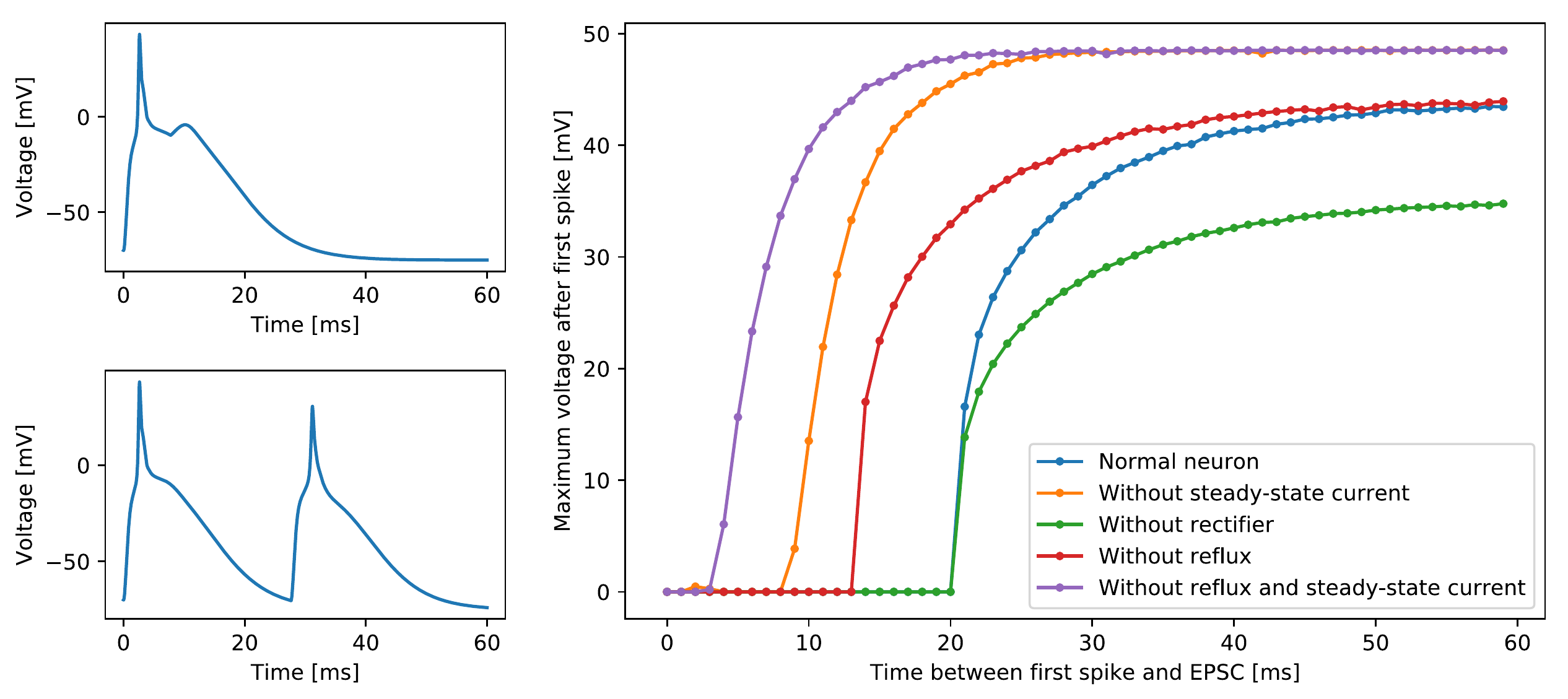}
        \put(0,42){A}
        \put(0,20){B}
        \put(34,42){C}
    \end{overpic}

    \caption[Refractory]{\textbf{Excitability of an MNN neuron after spiking.} 
    (\textit{A,B}) Voltage of an example neuron receiving two identical EPSCs (\textit{A}) $7~\text{ms}$ apart and (\textit{B}) $25~\text{ms}$ apart.
    (\textit{C}) Maximum voltage reached in response to the second EPSC for different time lags between the inputs. The first EPSC always generates a spike. The abscissa displays the time differences between its peak and the onset of the second EPSC. The ordinate displays the highest voltage reached after the end of the first spike, defined as reaching $0~\text{mV}$ from above. A plotted value of $0~\text{mV}$ means that the neuron did not exceed $0~\text{mV}$ after its first spike. 
    }
    \label{fig:refractory}
\end{figure}

As a single AP evokes an AP in a resting postsynaptic neuron and synapses are bidirectional, one might expect that the postsynaptic AP (or even the reflux) in turn evokes further presynaptic APs. However, experiments in two-neuron systems do not observe such repetitive firing but only bumps of depolarization after a spike \citep{Anderson_1985}. This is likely due to the long refractory period of scyphozoan neurons, which is initially absolute for about $30~\text{ms}$ and thereafter relative for about $70~\text{ms}$ \cite{Anderson_Schwab_1983}. In agreement with experimental findings, we do not observe repetitive firing in systems of two synaptically connected model neurons, but only bumps of depolarization after a spike. This indicates that our model neurons have a sufficiently long refractory period, although it has not been explicitly inserted. Fig.~\ref{fig:refractory} A shows as an example the voltage trace of a neuron that is stimulated by an EPSC, spikes and receives an EPSC due to the spiking of a postsynaptic neuron. Due to signal transmission delays, the neuron receives the second EPSC $7~\text{ms}$ after the first one.

To determine the refractory period effective under arrival of synaptic inputs, we apply two EPSCs with increasing temporal distance (see Fig.~\ref{fig:refractory} C). We find a refractory period of about $20~\text{ms}$. The longer refractory periods observed in scyphozoan neurons may be due to additional channel features that are not detectable from the voltage clamp data, such as delayed recovery from inactivation \citep{Kuo_Bean_1994, French_Zeng_Williams_Hill-Yardin_OBrien_2016}. The synaptic and AP traveling delay in our model (at most $3.5~\text{ms}$, see Methods) plus the time to reach threshold (about $2.5~\text{ms}$) are far from sufficient for the presynaptic neuron to recover from its spike, such that repetitive spiking is prevented, as observed in experiments.

To understand the origin of the effective refractory period's long duration, we determine it also in deficient model neurons, where the slow steady-state channel, the synaptic reflux and/or the synaptic rectifier \citep{Anderson_1985} are missing (Fig.~\ref{fig:refractory} C). We find that the synaptic reflux and the steady-state current are crucial for the long duration: without them the refractory period is reduced to about $5~\text{ms}$ (purple trace in Fig.~\ref{fig:refractory} C). In contrast, deactivation of the synaptic rectifier does not shorten the refractory period, but reduces the amplitude of the action potential, since the reversal potential of the channels is $+4~\text{mV}$. The synaptic rectifier thus allows spike peaks to more clearly exceed the $+20~\text{mV}$ threshold for synaptic transmission activation. It may therefore increase the reliability of signal conduction in the MNN.

\subsection{Modeling the Motor Nerve Net}
\subsubsection*{Qualitative Dynamics}
Given the described qualitative properties of its neurons and synapses, we can explain the main feature of the MNN, namely throughconductance without pathological firing: In fact, the properties of the MNN indicate that during the activation wave following an arbitrary initial stimulation of the network, every neuron spikes exactly once. Generally, this is the case in a network where (i) the synapses are bidirectional, (ii) a presynaptic action potential evokes action potentials in all non-refractory postsynaptic neurons and (iii) the refractory period is so long that there is no repetitive firing in two neuron systems.

This becomes clear if we think of the nerve net as a connected undirected graph with neuron dynamics evolving in discrete time steps. The undirectedness of the graph reflects the synaptic bidirectionality, point (i) above. We assume that it takes a neuron one time step to generate an AP; its postsynaptic neurons that are resting generate an AP in the next time step, see point (ii). After an AP, a neuron is refractory for at least one time step and thereafter becomes resting, ensuring (iii). More formally speaking, each vertex can be in one of three states in any time step: resting, firing, refractory. The state dynamics obey the following rules:
\begin{enumerate}
    \item If a vertex is firing at time step $t_i$, every connected, resting vertex will fire at $t_{i+1}$.
    \item If a vertex is firing at $t_i$, it will be refractory at $t_{i+1}$.
    \item If a vertex is refractory at  $t_i$, it will be resting at  $t_{i+1}$.
\end{enumerate}
 If in such a graph a number of vertices fires at  $t_0$ while the other vertices are resting (initial stimulation), every vertex will subsequently fire exactly once:
 Obviously any vertex $X$ will be firing at  $t_x$, where $x$ is the minimum of the shortest path lengths to any of the vertices firing at $t_0$.
Further, if a vertex $Y$ is firing at  $t_y$, where $y = x+s$, there must be a vertex $X$ firing at time $t_x$ with a path from $X$ to $Y$ with path length $s$. We will now assume that a vertex $X$ is not only firing at $t_x$ but also at $t_{x'}$ and show that this is impossible as it leads to a contradiction: We have $x' > x$ since $t_x$ is by definition the first time that $X$ fires after the initial stimulation.
Since the vertex is refractory at $t_{x+1}$ and resting at $t_{x+2}$, even $x' > x+2$ holds.
Let $x' = x+j$ where $j>2$. This implies that at $t_x$ a vertex $Y$ must be firing, with a path between $X$ and $Y$ of length $j$, along which the firing spreads from $Y$ towards $X$. There is, however, also a chain of firing traveling along this path from $X$ to $Y$.
If $j$ is even this results in two vertices in the center of the path firing right next to each other at $t_{x+\frac{j}{2}}$. After that both vertices are refractory and no other vertex along this path is firing.
If $j$ is odd there are two vertices firing at  $t_{x+ \frac{j-1}{2}}$ with a single vertex separating them. This vertex fires in the next time step, but since both neighboring vertices on this path are then refractory, no vertex along this path fires after that. Both cases contradict the initial assumption that $X$ spikes at $t_{x+j}$. We may thus conclude that $X$ fires only once.

\subsubsection*{Geometry}
To model the MNN in more detail, we uniformly distribute the developed Hodgkin-Huxley type neurons on a disc representing the subumbrella of a jellyfish with diameter $4~\text{cm}$. Its margin and a central disc are left void to account for margin and manubrium (see Methods for further details).  Eight rhopalia are regularly placed at the inner edge of the margin. We model their pacemakers as neurons which we stimulate via EPSCs to simulate a pacemaker firing. The neurons are geometrically represented by their neurites, modeled as straight lines of length $5~\text{mm}$ \citep{HORRIDGE_1954}. At the intersections of these lines lie connecting synapses \citep{Anderson_1985, Anderson_Gruenert_1988}. All synapses are bidirectional and have the same strength, sufficient to evoke an AP in a postsynaptic neuron. We incorporate neurite geometry and relative position into our single compartment models by assuming that the delay between a presynaptic spike and the postsynaptic EPSP onset is given by the sum of (i) the traveling time of the AP from soma to synapse on the presynaptic side, (ii) the synaptic transmission delay and (iii) the traveling time of the EPSC from synapse to soma on the postsynaptic side. The traveling times depend linearly on the distances between synapse and somata; for simplicity, we assume that AP and EPSC propagation speeds are equal. In agreement with \cite{Anderson_1985}, the total delays vary between $0.5~\text{ms}$ and $1.5~\text{ms}$.

Interestingly, the preferred spatial orientations of MNN neurites along the subumbrella are related to neuron position. \cite{HORRIDGE_1954} reports the following observations: 
\begin{itemize}
    \item Near the rhopalia, most neurites run radially with respect to the jellyfish center.
    \item Near the outer bell margin and between two rhopalia, most neurites follow the edge of the bell. 
    \item Closer to the center of the subumbrella there is no obvious preferred direction.
\end{itemize}
To incorporate these observations, we draw the neurite directions from distributions whose mean and variance depend appropriately on neuron position. Specifically, we use von Mises distributions for the angle, which are a mathematically simple approximation of the wrapped normal distribution around a circle \citep{Mardia_Jupp_1999}.

The neurite orientation structure may emerge due to ontogenetic factors: In the complex life cycle of scyphozoans, juvenile jellyfish start to swim actively during the ephyra stage. In this stage the jellyfish has some visual similarity to a starfish, with a disc in the center containing the manubrium, and eight (or more) arms, one per rhopalium, extending from it. The motor nerve net is already present in the ephyra and extends into its arms \citep{Nakanishi_Hartenstein_Jacobs_2009}. As the jellyfish matures, the arms grow in width until they fuse together to form the bell. MNN neurites simply following the directions of growth would thus generate a pattern as described above: Neurites in the center disc may not have a growth direction or constraints to follow, therefore there is no preferred direction. When the  ephyral arms grow out, neurites following the direction of growth run radially. Also the geometric constraints allow only for this direction. Neurons that develop in new tissue as the arms grow in width to form the bell, orient circularly, following the direction of growth.

\subsubsection*{Network Statistics}
\begin{figure}[htb!]
    \begin{fullwidth}

    \begin{overpic}[width=.9\paperwidth]{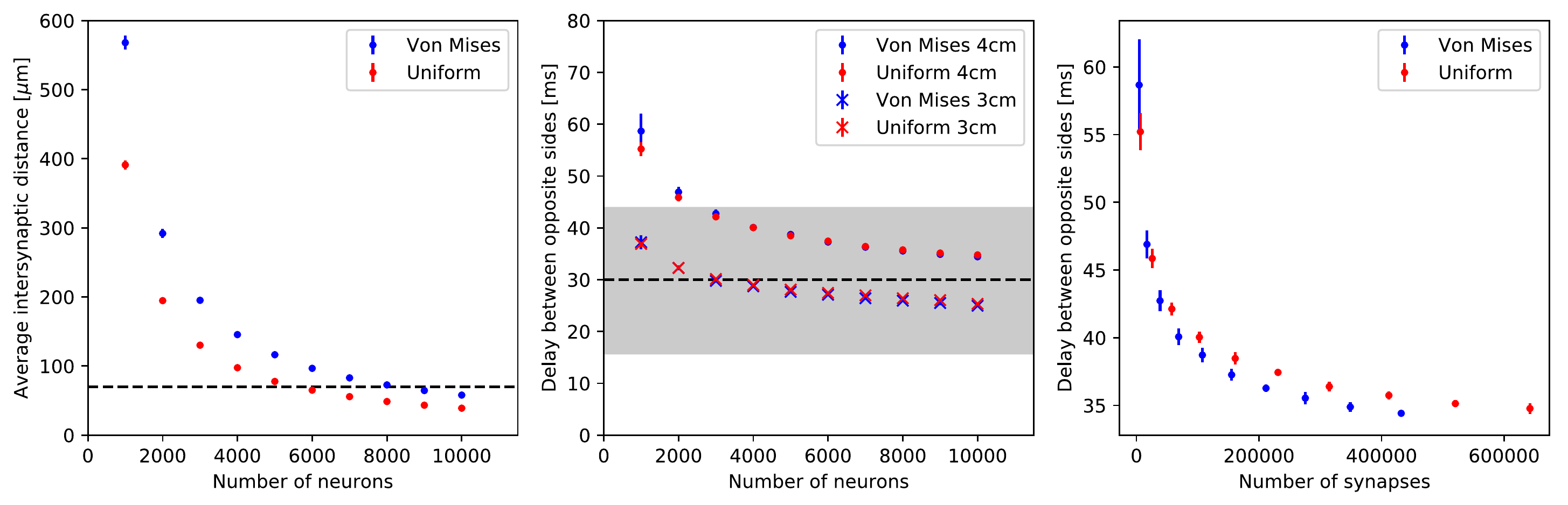}
        \put(0,30){A}
        \put(34,30){B}
        \put(67,30){C}
    \end{overpic}

    \caption[Number of Neurons]{\textbf{Synaptic density and activity propagation speed in von Mises and uniform MNNs.} 
    (\textit{A}) Average intersynaptic distance as a function of neuron number in von Mises and uniform MNNs. The dashed line indicates $70~\text{\textmu m}$ \citep{Anderson_1985}. (\textit{B}) Delay between the spiketimes of the pacemaker initiating an activation wave and the opposing one, for different MNN neuron numbers. Displayed are results for model jellyfish with $3~\text{cm}$ and $4~\text{cm}$ diameter. The dashed line indicates the experimentally measured average delay of $30~\text{ms}$ between muscle contractions on the initiating and the opposite side of \textit{Aurelia aurita} \citep{Gemmell_Troolin_Costello_Colin_Satterlie_2015}; the gray area shows its $\pm 1~$std.~dev. interval. (\textit{C}) Delays measured in (B) for the $4~\text{cm}$ jellyfish, plotted against the average number of synapses in MNNs with identical size.
    Measurement points are averages over 10 MNN realizations; bars indicate one standard deviation. 
    }\label{fig:NeuronNumbers}
    \end{fullwidth}
\end{figure}

There are, to our knowledge, no estimates on the number of neurons in a scyphozoan MNN; only some measurements for hydrozoans and cubozoans exist \citep{Bode_Berking_David_Gierer_Schaller_Trenkner_1973,Garm_Poussart_Parkefelt_Ekstrom_Nilsson_2007}.
However, \cite{Anderson_1985} measured the synaptic density in the MNN of \textit{Cyanea capillata}: the average distance between two synapses along a neurite is approximately $70~\text{\textmu m}$. For a neuron of $5~\text{mm}$ length this translates to roughly 70 synapses placed along its neurites. To obtain an estimate for the number of MNN neurons from this, we generate model networks with different neuron numbers, calculate their average synaptic distances and compare them with the experimentally observed values (see Fig.~\ref{fig:NeuronNumbers} A). We find that in a von Mises MNN, about 8000 neurons yield the experimentally measured synaptic density, while the uniform MNN requires about 5000 neurons. In general for a fixed number of neurons a von Mises MNN is more sparsely connected than a uniform MNN:
The biased neurite direction at the bell margin of a von Mises MNN (see Fig. \ref{fig:Networks}) implies that neurons in close proximity have a high probability of possessing similarly oriented neurites. This decreases their chance of overlap and thus the number of synapses.

\subsubsection*{Waves of Activation in the MNN}
\begin{figure}[htb!]
    \centering
     \begin{overpic}[width=\textwidth]{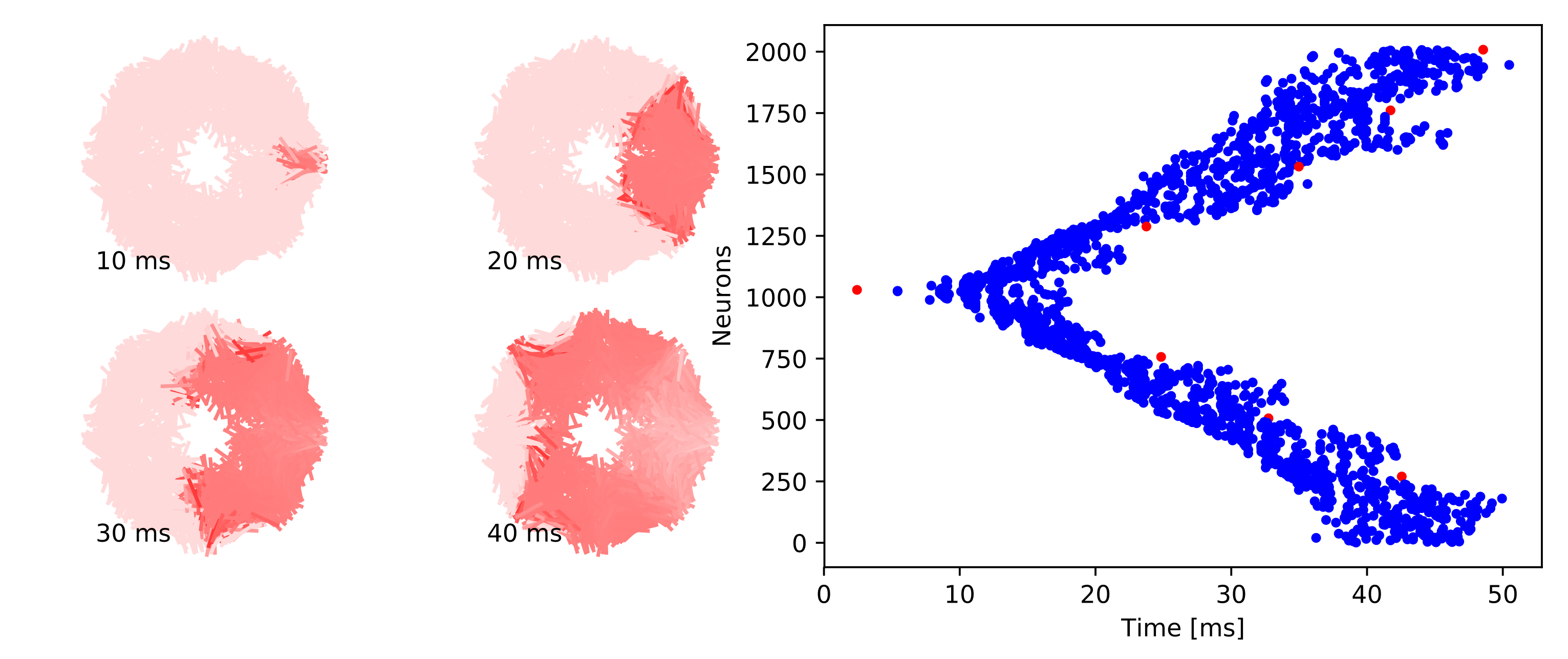}
        \put(0,40){A}
        \put(45,40){B}
    \end{overpic}
    \caption[2000 Neurons]{\textbf{Wave of activation in a von Mises MNN with 2000 neurons. } 
     (\textit{A}) Activity of each neuron at different times after stimulation of a single pacemaker neuron. Color intensity increases linearly with neuron voltage. (\textit{B}) Spike times of the same network. Neurons are numbered by their position on the bell. Red dots represent the pacemakers inside one of the eight rhopalia. The neurite orientations are distributed according to location-dependent von Mises distributions.
    }
    
    \label{fig:2000vM}
\end{figure}

\begin{figure}[htb!]
    \centering
     \begin{overpic}[width=\textwidth]{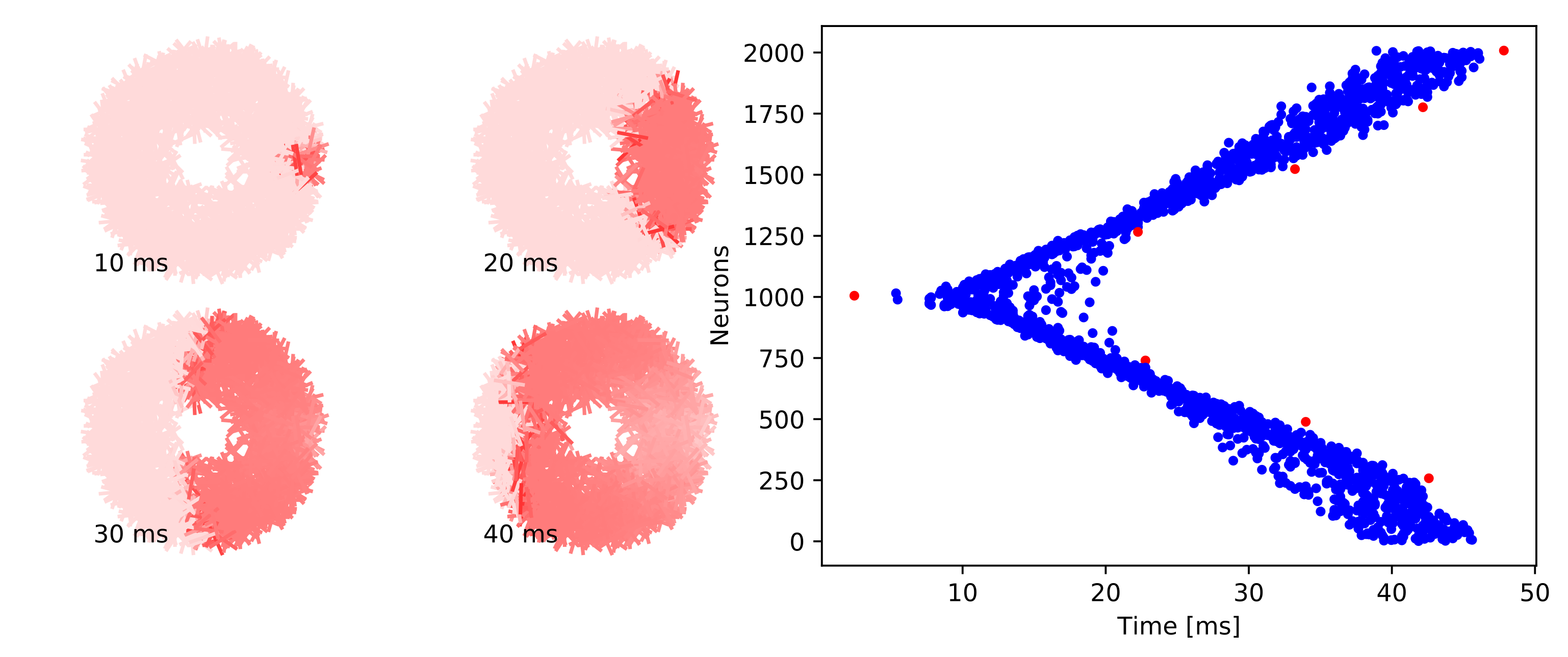}
        \put(0,40){A}
        \put(45,40){B}
    \end{overpic}
    \caption[2000 R Neurons]{\textbf{Wave of activation in a uniform MNN with 2000 neurons. } 
    Setup similar to Fig. \ref{fig:2000vM}, but the neurite orientations are uniformly distributed.
    }
    
    \label{fig:2000random}
\end{figure}

\noindent
Our numerical simulations confirm that firing of a pacemaker initiates a wave of activation where every MNN neuron generates exactly one AP (see Figs.~\ref{fig:2000vM}, \ref{fig:2000random} for an illustration). The activity propagates in two branches around the bell. These cancel each other on the opposite side. During the wave, all other pacemakers fire as well, which presumably resets them in real jellyfish. In a uniform MNN the wave spreads rather uniformly (Fig.~\ref{fig:2000random}). In a von Mises MNN the signal travels fastest around the center of the jellyfish and spreads from there, sometimes traveling a little backwards before extinguishing (Fig.~\ref{fig:2000vM}).

\cite{Gemmell_Troolin_Costello_Colin_Satterlie_2015} observed a delay between the muscle contractions on the initiating and the opposite side of about $30~\text{ms}$ (std.\ dev.\ $14~\text{ms}$), in \textit{Aurelia aurita} of $3-4~\text{cm}$ diameter. This delay should directly relate to the propagation of neural activity. We thus compare it to the delay between spiking of the initiating pacemaker and the opposing one in our model MNNs. We find that both our von Mises and uniform MNNs can generate delays within one standard deviation of the measurements, see Fig.~\ref{fig:NeuronNumbers} B. Our simulations indicate that MNN networks typically have 4000 neurons or more, as the propagation delays obtained for jellyfish with 3 and 4 cm diameter start to clearly bracket the experimentally found average at this size.

Fig.~\ref{fig:NeuronNumbers} B shows that the delay decreases with neuron density. On the one hand, this is because in denser networks among the more synaptic partners of a neuron there will be some with better positions for fast wave propagation; in other words, the fastest path from the initiating pacemaker to the opposing one will be better approximated, if the neurons have more synaptic partners to which the activity propagates. On the other hand, there is a decrease of delay due to stronger stimulation of neurons in denser networks: a postsynaptic neuron fires earlier if more presynaptic neurons have fired, since their EPSCs add up.

Both von Mises and uniform MNNs reach similar propagation speeds with the same number of neurons (Fig.~\ref{fig:NeuronNumbers} B), but von Mises MNNs have fewer synapses (Fig.~\ref{fig:NeuronNumbers} C). This implies that von Mises MNNs create more optimal paths of conduction. Indeed, neurons near the pacemaker preferably orient themselves radially towards the center of the subumbrella, and thus quickly direct the activity towards the opposite side. Since transmitter release consumes a significant amount of energy \citep{Niven_2016}, we conclude that von Mises networks are more efficient for fast throughconduction than uniform ones.


\subsection{A Model of Straight Swimming}
\begin{figure}[htb!]
    \begin{fullwidth}
    \centering
    \begin{overpic}[width=.28\paperwidth]{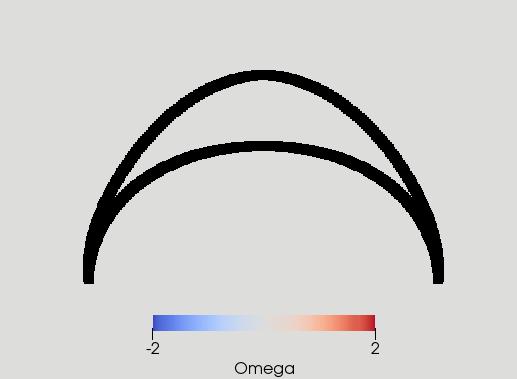} 
        \put(5,65){0.0 s}
    \end{overpic}
    \begin{overpic}[width=.28\paperwidth]{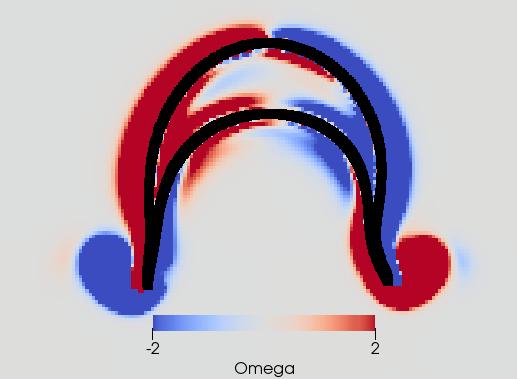} 
        \put(5,65){0.2 s}
    \end{overpic}
    \begin{overpic}[width=.28\paperwidth]{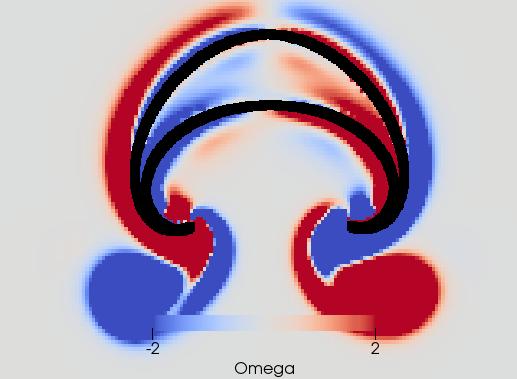} 
        \put(5,65){0.4 s}
    \end{overpic}\\
    \begin{overpic}[width=.28\paperwidth]{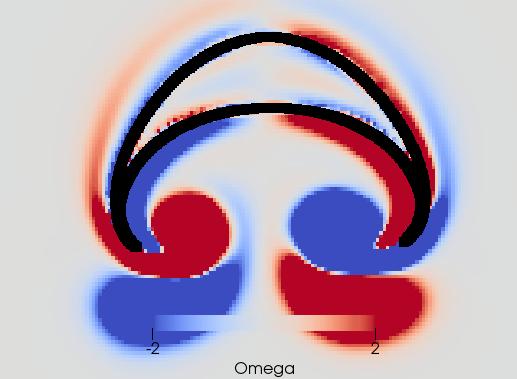} 
        \put(5,65){0.6 s}
    \end{overpic}
    \begin{overpic}[width=.28\paperwidth]{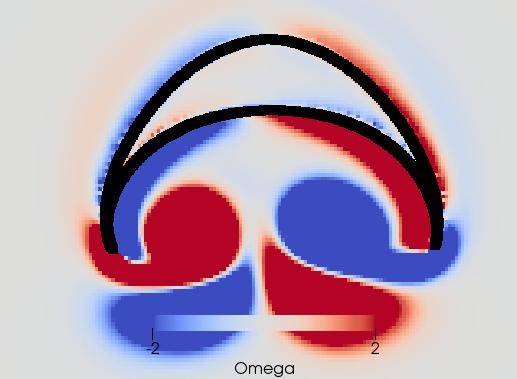} 
        \put(5,65){0.8 s}
    \end{overpic}
    \begin{overpic}[width=.28\paperwidth]{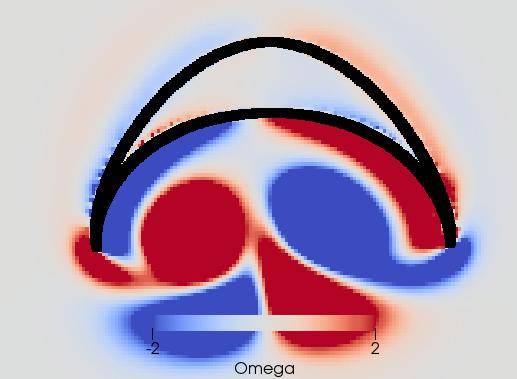} 
        \put(5,65){1.0 s}
    \end{overpic}
    \caption[Swimming]{\textbf{Swimming stroke evoked by a wave of activation in the MNN.} The panels show the dynamics of the bell surface (black) and internal and surrounding media (grey), in steps of $200~\text{ms}$. Coloring indicates medium vorticity $\Omega$ (in $1/\text{s}$, red a clockwise eddy and blue an anticlockwise one. In this and all following figures, it is the pacemaker on the left hand side of the bell that initiates MNN activation. Further, if not stated otherwise, the MNN has 10000 neurons.
    }
    
    \label{fig:Swimming}
    \end{fullwidth}
\end{figure}
\subsubsection*{MNN Activation and Swimming Strokes}
To analyze the swimming behavior, we employ a 2D hydrodynamics simulation of a cross section of the jellyfish bell. We assume that MNN neurons synaptically connect to muscles that lie in the same region (see Methods for details). APs in the neurons evoke twitches of the muscles. These add up to muscle contraction forces. Their interaction with the elastic forces of the bell and the hydrodynamics of the media in- and outside the bell determines the dynamics of the swimming stroke. Fig.~\ref{fig:Swimming} shows a representative time series of such a stroke. The left hand side pacemaker initiates a wave of MNN activation, which in turn triggers a wave of contraction around the subumbrella. Because the MNN activation wave is fast compared to muscle contraction and swimming movement, the motion is highly symmetrical. As a result, the jellyfish hardly turns within a stroke.

We can qualitatively compare the simulated swimming motion to that of real jellyfish by considering the formation of vortex rings. Earlier research suggests that the formation of two vortex rings pushes oblate jellyfish, such as \textit{Aurelia}, forward \citep{Dabiri_Colin_Costello_Gharib_2005, Gemmell_Costello_Colin_Stewart_Dabiri_Tafti_Priya_2013, Gemmell_Troolin_Costello_Colin_Satterlie_2015}. In a 2D cross section, a vortex ring is reflected by a vortex pair with opposing spin. We find indeed that two such vortex pairs are shed off near the bell margin (see Fig.~\ref{fig:Swimming}). The first pair is shed off during the contraction and the second one during the relaxation. The second pair slips under the jellyfish bell, which provides additional forward push \citep{Gemmell_Costello_Colin_Stewart_Dabiri_Tafti_Priya_2013}.

\begin{figure}[htbp!]
    \centering
     \begin{overpic}[width=0.8\textwidth]{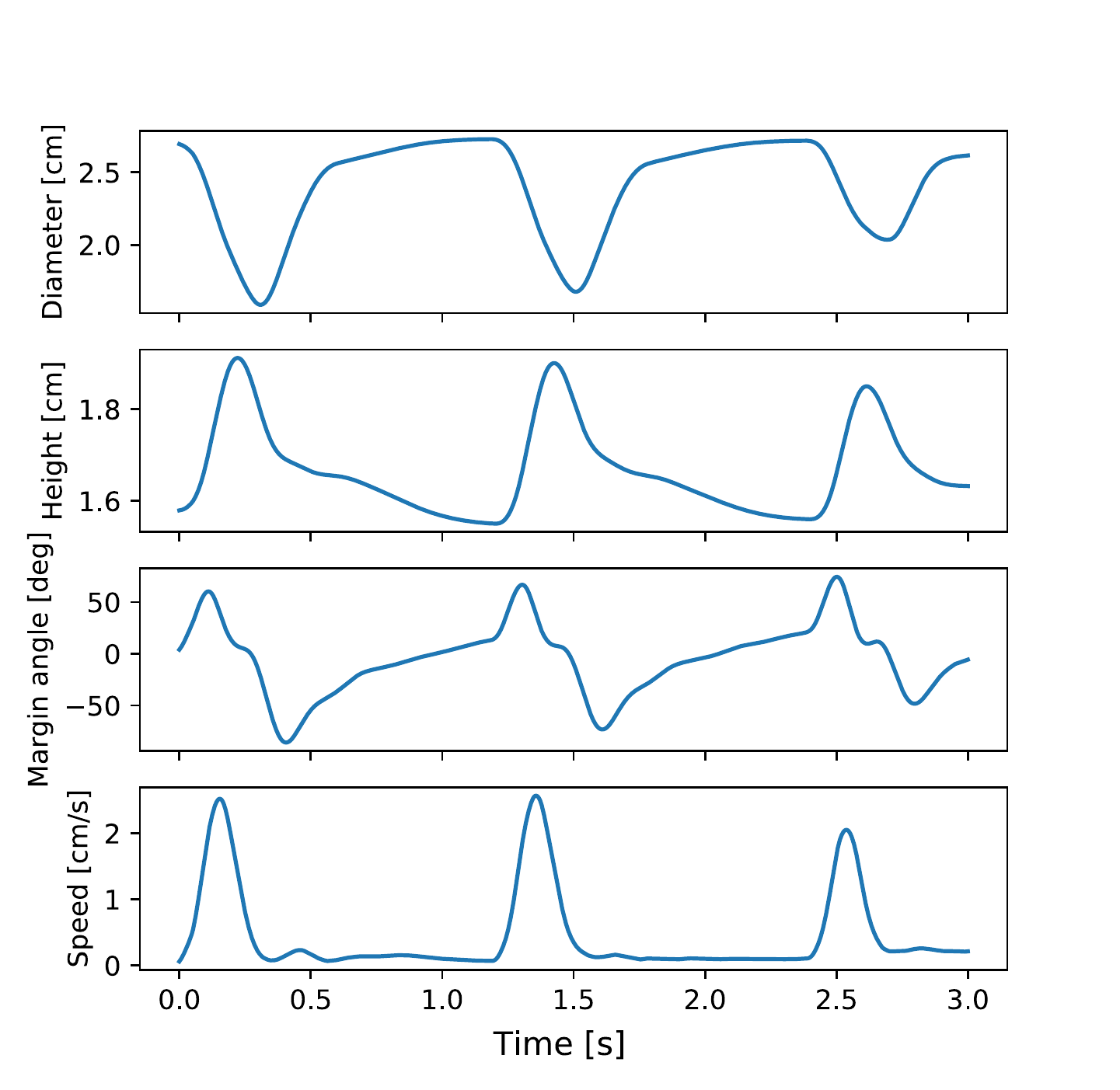}
        \put(-2,85){A}
        \put(-2,65){B}
        \put(-2,45){C}
        \put(-2,25){D}
    \end{overpic}
    \caption[McHenry]{\textbf{Characteristics of bell shape during swimming.} Dynamics of \textit{(A)} bell diameter, \textit{(B)} bell height and \textit{(C)} the orientation of the margin of the bell relative to the orientation of the bell as a whole, during a sequence of swimming strokes as in Fig.~\ref{fig:Swimming} A, initialized in intervals of 1.2 s. \textit{(D)} Corresponding speed profile. All values are measured as in the model by \cite{McHenry_Jed_2003}.
    }
    
    \label{fig:McHenry}
\end{figure}

\cite{McHenry_Jed_2003} measured changes in the bell geometry of \textit{Aurelia aurita} during its swimming motion. When tracking the same data in our simulations we find qualitatively similar time series. In particular, the sequence of changes in the bell geometry agrees with that of real jellyfish: During the contraction phase the bell diameter shrinks and the bell height increases. The bell margin begins to bend outward as the jellyfish contracts and folds inward during the relaxation of the bell. Furthermore, we can adapt the parameters of our model to achieve a quantitative agreement of the bell characteristics and the reached speeds (see Fig.~\ref{fig:McHenry}, \cite{McHenry_Jed_2003} Fig.~2).
 The experiments, however, show broader speed peaks and a longer continuation of forward movement after bell relaxation. This may be due to differences in vortex dynamics in 2D and 3D, see Discussion.

\begin{figure}[htb!]
    \centering
     \begin{overpic}[width=\textwidth]{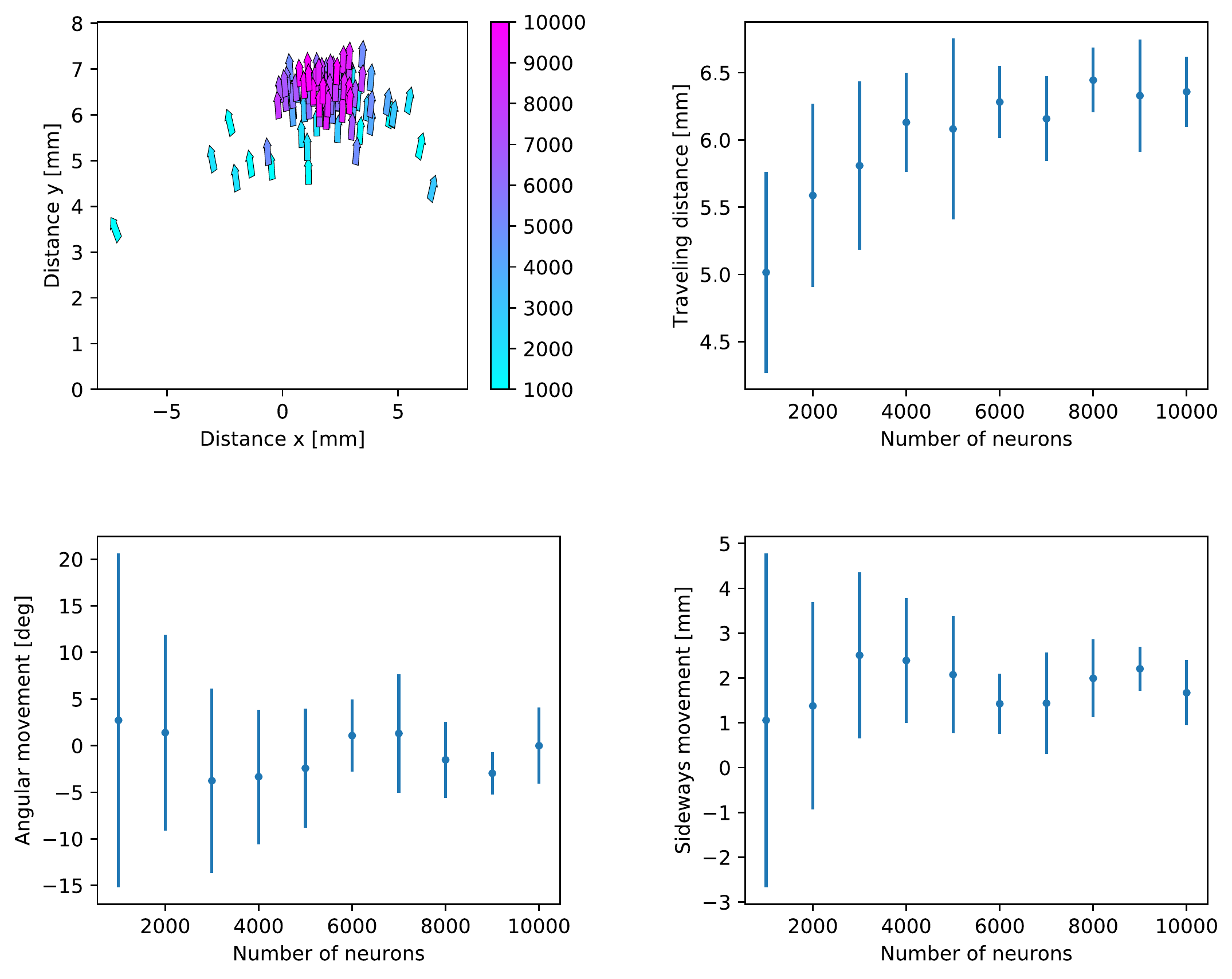}
        \put(0,77){A}
        \put(49,77){B}
        \put(0,35){C}
        \put(49,35){D}
    \end{overpic}
    \caption[Swimstats]{\textbf{Characteristics of swimming strokes for different MNN sizes.} (\textit{A}) shows the distance traveled within a single swimming stroke (origins of arrows) and the orientation after the stroke (direction of arrows) for 100 jellyfish with different MNNs. Color indicates the MNN sizes, which range in 10 steps from 1000 to 10000. (\textit{B,C,D}) vizualize the dependence of the distributions of swimming characteristics on MNN size. (\textit{B}) shows the total distances traveled, (\textit{C}) the angular movements (i.e.~angular changes in spatial orientation, in degrees) and (\textit{D}) the distances moved perpendicularly to the original orientation of the jellyfish. Measurement points are the averages of the 10 jellyfish with MNNs of the same size in (A), bars indicate one standard deviation.
    }
    \label{fig:Swimstat}
\end{figure}
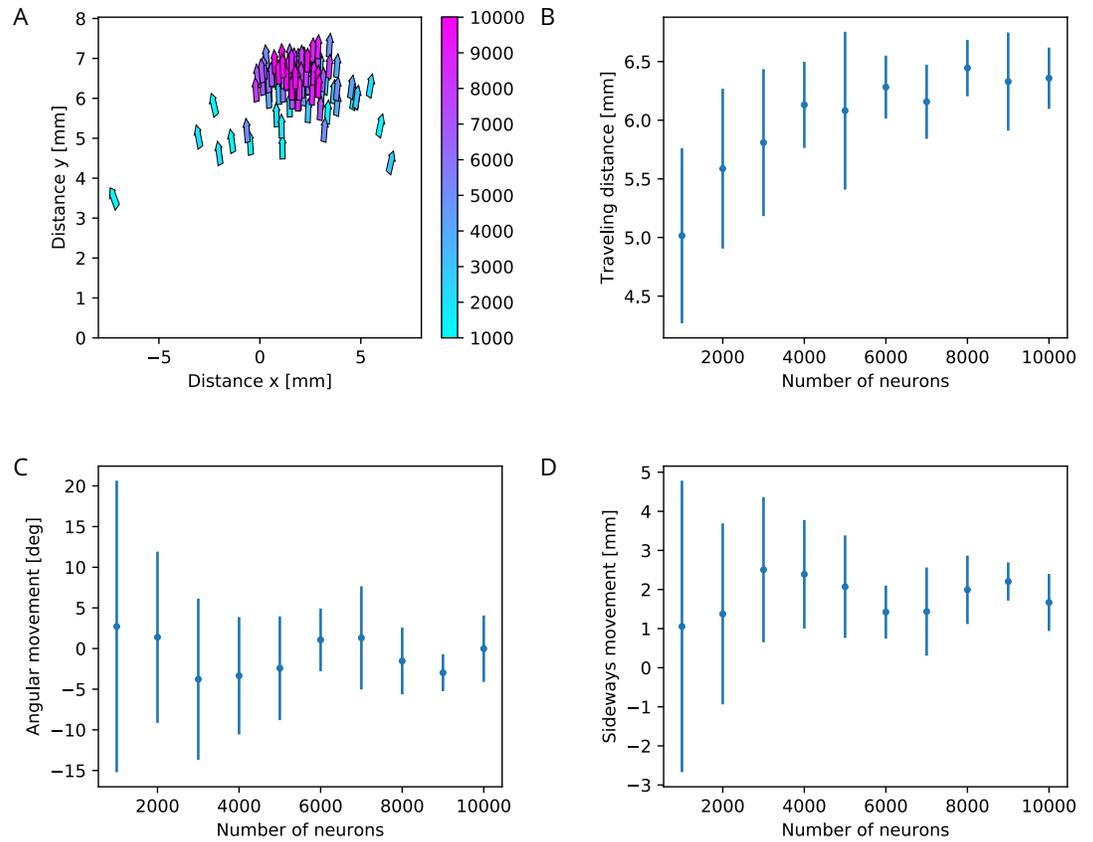

\subsubsection*{Influence of Network Size}
To quantify the effects of MNN size on swimming, we evaluate travel distances and changes in orientation, see Fig.~\ref{fig:Swimstat}. 
We find that the typical total distance traveled by individual jellyfish increases with network size (Fig.~\ref{fig:Swimstat} A,B), while the variance and thus the typical distance traveled sideways and the typical angular movement decrease (Fig.~\ref{fig:Swimstat} A,C and A,D). 
This can be explained by the higher temporal and spatial coherence in the activation waves of larger MNNs. They arise from larger throughconductance speed, see Fig.~\ref{fig:NeuronNumbers}, and from more uniform neuron density and muscle innervation: Since neurons are random uniformly distributed in space, the fluctuation of local neuron density relative to its mean decreases with increasing neuron number. This implies that the relative fluctuation in the number of neurons innervating the different muscle segments decreases. With small MNNs, random fluctuations in the number of innervating neurons are likely to lead to a spatial imbalance of contraction force that is sufficient to generate marked sideways movement and turning. Generally the variance of a characteristic sampled over different MNN realizations decreases as the number of neurons increases, because the decrease of relative local density fluctuations implies that the network ensembles become more homogeneous.

\begin{figure}[htb!]
    \begin{fullwidth}
    \centering
    \begin{overpic}[width=.28\paperwidth]{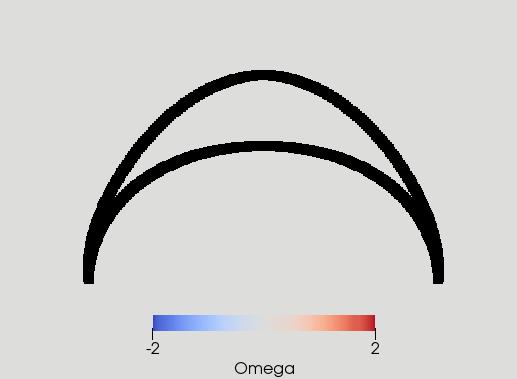} 
        \put(5,65){0.0 s}
    \end{overpic}
    \begin{overpic}[width=.28\paperwidth]{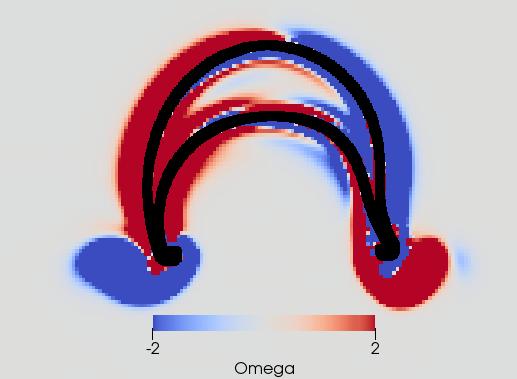} 
        \put(5,65){0.2 s}
    \end{overpic}
    \begin{overpic}[width=.28\paperwidth]{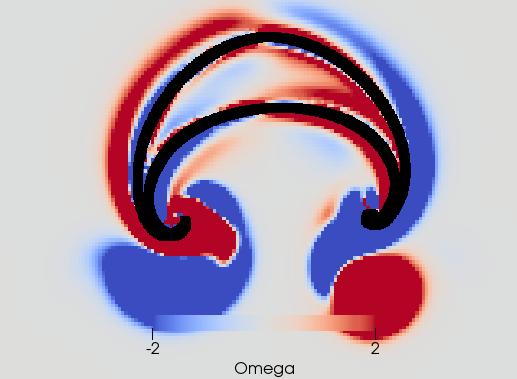} 
        \put(5,65){0.4 s}
    \end{overpic}\\
    \begin{overpic}[width=.28\paperwidth]{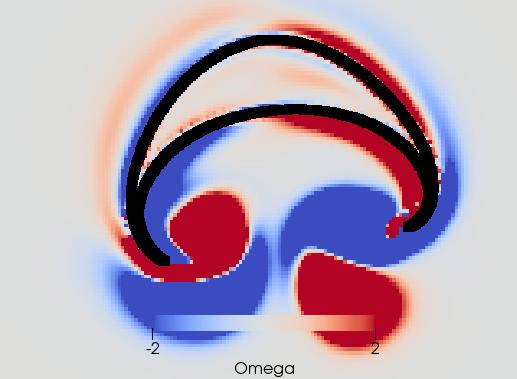} 
        \put(5,65){0.6 s}
    \end{overpic}
    \begin{overpic}[width=.28\paperwidth]{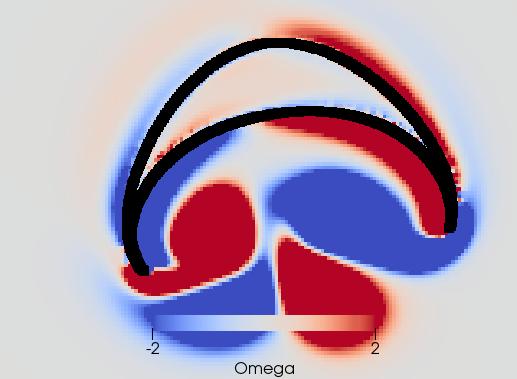} 
        \put(5,65){0.8 s}
    \end{overpic}
    \begin{overpic}[width=.28\paperwidth]{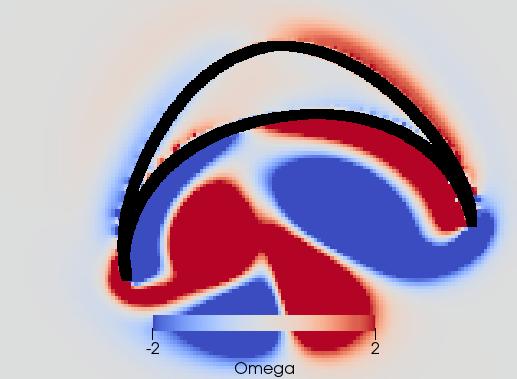} 
        \put(5,65){1.0 s}
    \end{overpic}
    \caption[Turning]{\textbf{Swimming stroke evoked by simultaneously initiated waves in MNN and DNN.} The activity in the DNN and MNN leads to a simultaneous contraction of the left bell margin and the left bell swimming muscles near the margin. The jellyfish therefore turns in the direction of the initiating rhopalium.
    The DNN has 4000 neurons. MNN and further description are as in Fig.~\ref{fig:Swimming}.
    }
    
    \label{fig:Turning}
    \end{fullwidth}
\end{figure}

\subsection{A Model of Turning}

\subsubsection*{The Mechanism of Turning}
Finally, we investigate whether the contraction of the bell margin due to DNN activity can lead to a turning mechanism similar to the one described by \cite{Gemmell_Troolin_Costello_Colin_Satterlie_2015}. In this study, radial muscles in the bell margin were found to contract on the inside of a turn, which was hypothesized to be controlled by DNN activation. Indeed, we find that a simultaneous activation of the DNN and the MNN leads to a turn, see Fig.~\ref{fig:Turning}. The jellyfish turns towards the origin of the contraction wave if both MNN and DNN are stimulated at the same time. The radial muscles of the bell margin on the stimulated side contract simultaneously with the circular muscles such that the bell margin stiffens up and does not bend outwards during the contraction of the bell, cf.~the left hand side margin in Fig.~\ref{fig:Turning}. Because the jellyfish has to displace more water on this side, the contraction is slowed down. Due to the different conduction speeds of MNN and DNN, the circular muscles on the other side contract before the radial muscles. The stroke is therefore similar to that during straight swimming, leads to a stronger contraction and turns the jellyfish towards the origin of the activation wave.

The displayed dynamics are similar to those experimentally observed in \textit{Aurelia} by \cite{Gemmell_Troolin_Costello_Colin_Satterlie_2015}. In particular, the jellyfish turns towards the side of initial contraction and the bell margin on the inside of the turn is contracted while the opposing one extends outwards. The margin bending in our model appears stronger than in \cite{Gemmell_Troolin_Costello_Colin_Satterlie_2015}. Further, the delay between the onsets of contraction on the initiating and the opposing sides is shorter in our model. Such dissimilarities may be brought into agreement by more detailed DNN and bell modeling in 3D hydrodynamic environments.

\subsubsection*{Relative Timing of MNN and DNN Activation}
\begin{figure}[htb!]
    \centering
    \begin{overpic}[width=\textwidth]{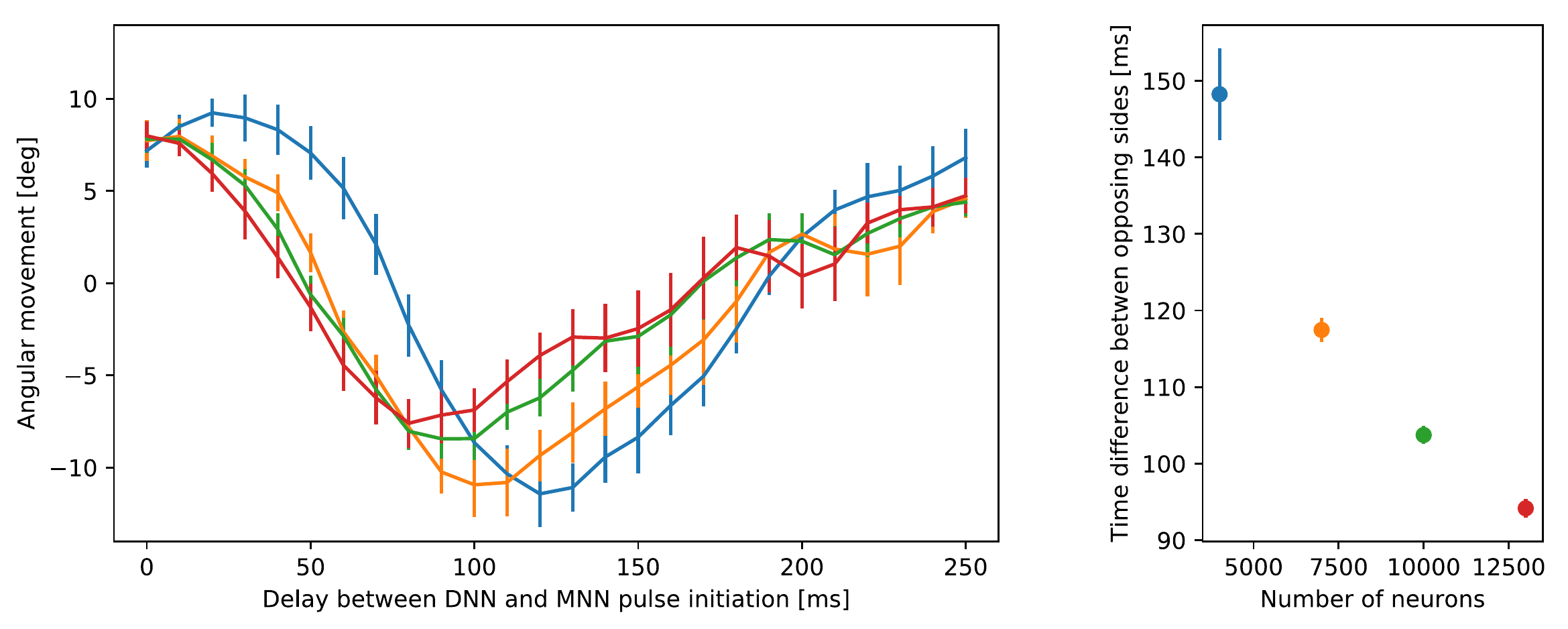}
        \put(0,37){A}
        \put(65,37){B}
    \end{overpic}
    \caption[Angle]{\textbf{Dependence of turning on the delay between DNN and MNN activation.} (\textit{A}) Angular movement of model jellyfish versus delay between DNN and MNN activation. The panel displays the angular movement one second after the initiation of the MNN. Turns towards the initiating rhopalium have positive angular movements, while turns away have negative ones. Blue, orange, green and red coloring indicates DNN sizes of 4000, 7000, 10000 and 13000 neurons.
    (\textit{B}) Delay between initiation of DNN activity and its reaching of the opposing side, as a function of the number of DNN neurons (similar to Fig.~\ref{fig:NeuronNumbers} B).
    Measurement points are averages over 10 realizations of MNNs with 10000 neurons and DNNs with the indicated size, bars indicate one standard deviation.
    }
    
    \label{fig:Angle}
\end{figure}

\begin{figure}[htb!]
    \begin{fullwidth}
    \centering
    \begin{overpic}[width=.28\paperwidth]{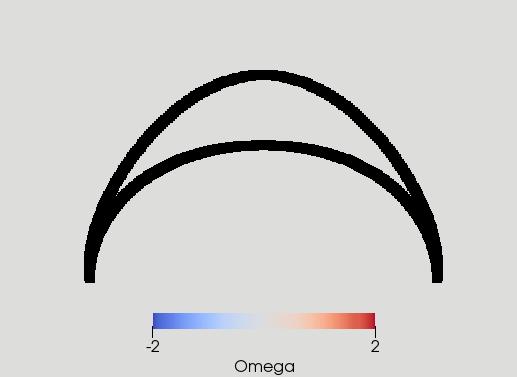} 
        \put(5,65){0.0 s}
    \end{overpic}
    \begin{overpic}[width=.28\paperwidth]{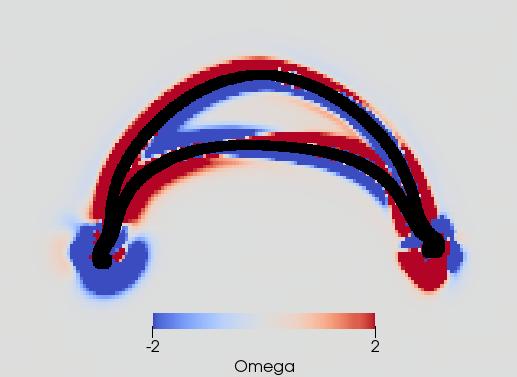} 
        \put(5,65){0.2 s}
    \end{overpic}
    \begin{overpic}[width=.28\paperwidth]{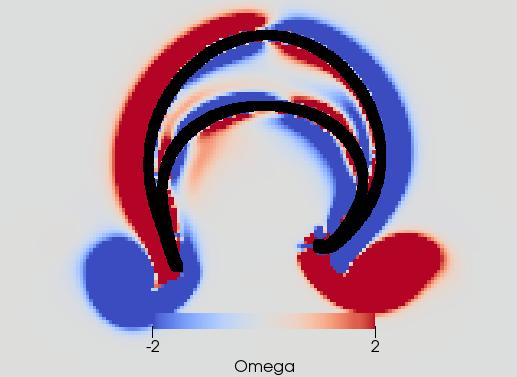} 
        \put(5,65){0.4 s}
    \end{overpic}\\
    \begin{overpic}[width=.28\paperwidth]{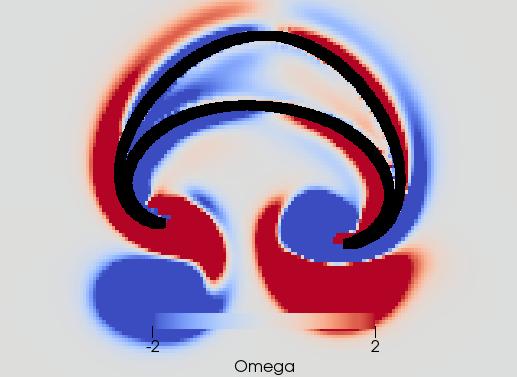} 
        \put(5,65){0.6 s}
    \end{overpic}
    \begin{overpic}[width=.28\paperwidth]{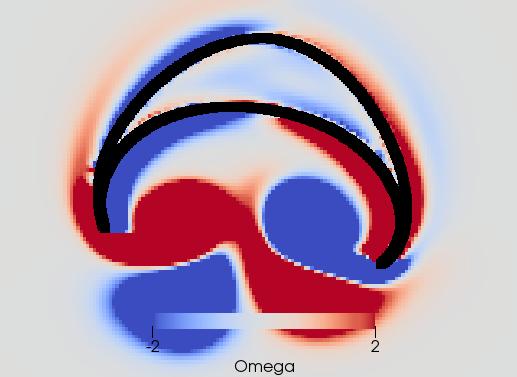} 
        \put(5,65){0.8 s}
    \end{overpic}
    \begin{overpic}[width=.28\paperwidth]{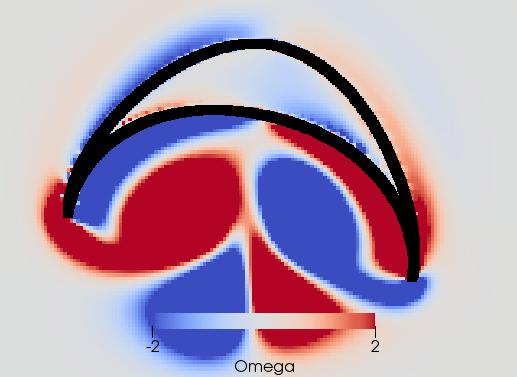} 
        \put(5,65){1.0 s}
    \end{overpic}
    \caption[Turning]{\textbf{Swimming stroke evoked by sequentially initiated waves in MNN and DNN.}
    Initiation of the MNN $120~\text{ms}$ after the DNN leads to a simultaneous contraction of the right bell margin and the right bell swimming muscles near the margin. The jellyfish therefore turns away from the direction of the initiating rhopalium. MNN and DNN as in Fig.~\ref{fig:Turning}.
    }
    \label{fig:OppositeTurn}
    \end{fullwidth}
\end{figure}
\cite{Passano_1965, Passano_1973} found that after externally stimulating the DNN, the MNN becomes active after a significant delay. We therefore study the impact of different delays between DNN and MNN activation on the turning behavior, see Fig.~\ref{fig:Angle} A. For small delays the jellyfish turns towards the origin of the stimulation, like for zero delay (Fig.~\ref{fig:Turning}) and as observed by \cite{Horridge_1956, Gemmell_Troolin_Costello_Colin_Satterlie_2015}. As the delay increases, the jellyfish turns less. At a certain delay the turning direction changes, and the jellyfish turns more and more into the opposite direction. For even larger delays the jellyfish again turns less and there is eventually another change of direction. The points of first direction change and maximum opposite turning depend on the speed of the DNN signal (Fig.~\ref{fig:Angle} B).

The first change of turning direction occurs because for sufficiently large delay between DNN and MNN the radial muscles on the side of wave initiation are already relaxing when the swimming muscles contract, see Fig.~\ref{fig:OppositeTurn}. On the opposing side the activity of the radial muscles then coincides with the contraction of the swimming muscles. Therefore the same mechanism that causes the turn towards the initiating rhopalium for simultaneous DNN and MNN activation lets the jellyfish now turn to the other side. This, although both DNN and MNN are activated by the same rhopalium. The most negative angular movement occurs at a delay that is about the conduction delay of the MNN shorter than the time it takes the DNN to conduct a signal around the bell; compare the delays at minima in Fig.~\ref{fig:Angle} A with the corresponding DNN conduction delays in Fig.~\ref{fig:Angle} B minus the MNN conduction delay of $35~\text{ms}$. With such a delay, the two signals will simultaneously reach the opposing side of the bell.

This previously undescribed mechanism may explain how a jellyfish is able to avoid undesired stimuli. After it is, for example, mechanically stimulated somewhere on its bell, the corresponding DNN excitation spreads and reaches the rhopalium closest to the origin of the stimulus. If the MNN would then fire immediately, the jellyfish would turn towards the stimulus. Our simulation together with the experiments by \cite{Passano_1965, Passano_1973} let us hypothesize that the pacemaker at the rhopalia may rather fire after an appropriate delay, generated by a yet unknown mechanism. This would allow the jellyfish to flee if necessary.

\section{Discussion}
We have built a multiscale model of the neuromuscular system of scyphozoan jellyfish on the basis of biophysical, physiological and anatomical data.  Our model reproduces known experimental findings and predicts new ones across multiple scales, from ion channel dynamics over neuron and neuronal network activity to animal behavior.

We propose a Hodgkin-Huxley-type neuron model for scyphozoan MNN neurons, on the basis of voltage-clamp data \citep{Anderson_1989}. The model yields an explanation for experimental findings, such as the long refractory period of MNN neurons \citep{Anderson_Schwab_1983}, in terms of ion channel and synapse dynamics. Furthermore, it makes experimentally testable predictions on the time course of different ion channel activations during an AP and the effect of their blocking. The number of parameters in the model could be reduced. For example the slow outward current does not contribute to the neuron dynamics in the considered physiological regime. It will be interesting to explore which parameter values are crucial for its functioning in the future.

We develop the idea of synaptic transmitter reflux as a natural consequence of the bidirectional synapses connecting MNN neurons \citep{Anderson_1985}. Our model indicates that the synaptic reflux generates a pecularity of the scyphozoan AP shape, namely a delayed decay or small voltage bump immediately after the return from peak AP depolarization, which is visible in experimental data \citep{Anderson_1985}.
Later voltage bumps occur since postsynaptic APs evoke EPSCs in the presynaptic neuron (Fig.~\ref{fig:refractory} A, B, \cite{Anderson_1985}, \cite{Anderson_Schwab_1983}).

A simple, phenomenological network model qualitatively incorporating key features of MNN neurons shows why MNN and DNN do not generate pathological activity, but a single wave of activation after an initial stimulation. The model predicts that during such a wave every neuron in the nerve net fires exactly once, no matter where the initial excitation originates.

We build a biologically more detailed neuronal network model of the scyphozoan MNN by placing the developed Hodgkin-Huxley-type neurons on a 2D geometry representing the subumbrella. Based on anatomical observations \citep{HORRIDGE_1954}, we propose that their neurite orientations are distributed according to location-dependent von Mises distributions. We study the dynamics of these von Mises MNNs and compare them to MNNs with location-independent random uniformly distributed neurite orientations. Similarly, we build a model for the DNN. Since no data on neurite orientation of the DNN is available, we assume a random uniform distribution.

Both our von Mises and uniform MNNs can reproduce the experimentally observed throughconduction delay of MNN activation waves. Von Mises MNNs are, however, more cost efficient in the sense that their waves require fewer synaptic transmitter releases to reach the same delay. The experimentally found biological features of the network structure thus provide a partial optimization compared to homogeneous random networks. We suggest that the structure may emerge in a simple manner as the neurites follow the directions of growth and geometric constraints during ontogenesis.

Our model suggests two estimates of the unknown number of neurons in a scyphozoan MNN. The first one is purely geometrical, based on our network structure and the average distance of synapses on neurites measured by \cite{Anderson_1985}. The second one accounts for the network dynamics and compares throughconduction delays in our models with experimentally measured ones in \cite{Gemmell_Troolin_Costello_Colin_Satterlie_2015}. The estimates indicate that the number of neurons is of the order of 10000 neurons in jellyfish of about $4~\text{cm}$ diameter. Possible error sources of the estimates include the mixing of data from animals of different species (\textit{Cyanea capillata} in \cite{Anderson_1985, Anderson_1989} and \textit{Aurelia aurita} in \cite{HORRIDGE_1954}) and sizes, distributed neurite lengths and the presence of multipolar cells and multiple synapses between neurons \citep{HORRIDGE_1954, Anderson_1985}. The obtained neuron numbers are within the range found for other cnidarians: hydrozoans and cubozoans have approximately 5,000 to 20,000 neurons \citep{Bode_Berking_David_Gierer_Schaller_Trenkner_1973, David_1973, Garm_Poussart_Parkefelt_Ekstrom_Nilsson_2007}.

To connect neural activity to behavior, we develop a model for the muscle system and the elastic bell of \textit{Aurelia aurita}. The MNN evokes the contractions of the swimming muscles. We place the resulting model jellyfish in a hydrodynamic environment and simulate its swimming behavior. To reduce the duration and complexity of the hydrodynamics simulations, we consider a 2D jellyfish model and environment. 
We observe shedding of vortex pairs in the surrounding medium and, after appropriately adjusting parameters of the fluid-structure simulation, bell geometry dynamics similar to experimental observations \citep{McHenry_Jed_2003,Dabiri_Colin_Costello_Gharib_2005, Gemmell_Costello_Colin_Stewart_Dabiri_Tafti_Priya_2013, Gemmell_Troolin_Costello_Colin_Satterlie_2015}.
The restriction to a 2D simulation setup entails limitations, at least for obtaining quantitatively accurate results: In 2D vortex pairs can move independently from one another, while 3D vortex rings move as one unit during real jellyfish swimming \citep{Dabiri_Colin_Costello_Gharib_2005}. Further, vortex rings in 3D expand while the corresponding vortex pairs with opposite vorticity in 2D approach each other. As a result, in our simulation the vortex pair released during the relaxation moves further into the jellyfish bell than a real vortex ring would. The difference in vortex dynamics may explain that our model jellyfish stops moving forward quickly during relaxation after a stroke in contrast to data (cf.~\cite{McHenry_Jed_2003,Gemmell_Costello_Colin_Stewart_Dabiri_Tafti_Priya_2013}). Other researchers found similar limitations when simulating oblate jellyfish in 2D \citep{Herschlag_Miller_2011}.
Previous work in 2D has only looked at a symmetric swimming motion, where vortex pairs are shed off perfectly symmetrically \citep{Rudolf_Mould_2009,Herschlag_Miller_2011, Gemmell_Costello_Colin_Stewart_Dabiri_Tafti_Priya_2013, Hoover_Miller_2015}. However, in our simulations, the contractions are slightly asymmetric, due to the throughconduction delay in the MNN. Since in 2D the resulting vortices move under the jellyfish bell and stay and accumulate there, they exert a strong asymmetric force after several swimming strokes initiated at the same rhopalium. 
To counteract this effect, we slightly increase the viscosity of the surrounding medium.
A simple model of a contraction wave with finite propagation speed has been tested in a 3D jellyfish simulation in \cite{Hoover_2015} and \cite{Hoover_Griffith_Miller_2017}.
They found that turning reduces with increasing propagation speed. We observe this as well when increasing the number of neurons in the MNN, which increases the propagation speed of its activation and the induced muscle contraction wave. A larger MNN also increases the distance traveled after each stroke, enhancing the swimming speed.

We find that the details of the muscle dynamics are not crucial for the effective swimming motion, since the contraction of the swimming muscles is explosive and does not result from the application of continuous tension (see recordings by \cite{Gemmell_Costello_Colin_Stewart_Dabiri_Tafti_Priya_2013}). Therefore, our model produces a swimming motion that appears realistic for a wide variety of parameters, with the restrictions discussed above.

Based on experimental findings \citep{Gemmell_Troolin_Costello_Colin_Satterlie_2015}, we incorporate radial muscles in the margin of our jellyfish model. They are activated by the DNN. If the DNN and the MNN are initiated at the same time by a rhopalium, they evoke a simultaneous contraction of the nearby bell margin and radial swimming muscles. Similar to the experimental observations, we find that this turns the jellyfish towards the initiation site \citep{Gemmell_Troolin_Costello_Colin_Satterlie_2015}. Such voluntary turning is large compared to involuntary turning during straight swimming strokes for the estimated number of MNN neurons.

After mechanical stimulation the DNN generates a wave of activation, which in turn initiates an MNN wave at the closest rhopalium \citep{Horridge_1956}. A turn towards this rhopalium and thus towards the site of stimulation may often be undesired. Our simulations indicate that appropriate delays between MNN and DNN activation induce turns away from the stimulation site. Strongest such turns occur for delays that let both excitation waves reach the opposite side at the same time. We hypothesize that the rhopalia generate appropriate delays and allow the jellyfish to avoid predators or crashing into obstacles \citep{Albert_2008}. This previously unknown level of control may be experimentally detected by measuring the timing of DNN and MNN activity, similar to \cite{Passano_1965}, while simultaneously recording the swimming motion of the jellyfish.

In our current model, the MNN and DNN are stimulated by an artificially induced spike in one of their neurons at the location of a rhopalium. For a more complete modeling of the nervous system, future research should develop a model for pacemakers and their activity. This requires further experiments on their response properties and sensory information integration \citep{Nakanishi_Hartenstein_Jacobs_2009,Garm_Ekstroem_Boudes_Nilsson_2006}. Such data will also be key to test our prediction of the jellyfish's ability to avoid predators or obstacles by turning away from them. This ability might, in addition to different timings of DNN and MNN activity, use some form of multisensory integration differentiating threats from harmless stimuli.

To conclude, in this study we built the first comprehensive model of the neuromuscular system of a cnidarian. Specifically we considered the jellyfish \textit{Aurelia aurita}. This is particularly relevant due to the position of jellyfish in the evolutionary tree and their highly efficient swimming motion. Our model reproduces experimental data on multiple scales and makes several experimentally testable predictions. The simulations suggest that the simple nerve net structure may be optimized to conduct signals across the bell. In addition we find that the nerve nets enable a higher level of turning control than previously thought to be present in a radially symmetric organism that only receives decentralized sensory information. Our study bridges the gap between single neuron activity and behavior in a comparatively simple model organism. It lays the foundation for a complete model of neural control in jellyfish and related species and indicates that such modeling approaches are feasible and fruitful. Our results may be particularly useful for creating models of ctenophores and cnidarians like \textit{Hydra vulgaris}, where observing the complete nervous system of a living animal is possible \citep{Dupre_Yuste_2017, szymanski2019mapping}. A comparative computational analysis of their different nervous system dynamics and behavior could shed light on the early evolution of nervous systems.

\section{Methods}
\subsection{Neuron Model}
\label{sec:Neuron}
We use the voltage-clamp and action potential data of \cite{Anderson_1989, Anderson_1985} to develop a biophysical single compartment model of a scyphozoan neuron. The model describes the dynamics of the neuron's membrane potential $V$ and its transmembrane currents.
Following \cite{Anderson_1989}, we incorporate a transient inward current ($I_{\text{I}}$) and three outward currents: a steady-state outward current ($I_{\text{SS}}$) and a slow and a fast transient outward current ($I_{\text{ST}}$ and $I_{\text{FT}}$, respectively). Furthermore, we include a passive leak current ($I_{\text{L}}$). The membrane voltage thus follows the ordinary differential equation
\begin{equation}
    C_m \frac{\mathrm{d} V}{\mathrm{d} t} = I_{\text{syn}} - I_{\text{I}} - I_{\text{FT}} - I_{\text{ST}} - I_{\text{SS}} - I_{\text{L}},
    \label{eq:model}
\end{equation}
where $C_m$ is the membrane capacitance and $I_{\text{syn}}$ the synaptic input current (see next section).
The currents are modeled with a Hodgkin-Huxley type gate model \citep{Izhikevich_2007}. The steady-state current has a single gating variable $G_g$; exponentiation with a suitable exponent ${p_g}$ yields the probability that an individual channel is open. Transient currents have two gating variables, one for activation and one for inactivation. For these currents, the probability that an individual channel is open is given by the product of the two gating variables after exponentiation with suitable exponents. The transmembrane currents are thus given by
\begin{subequations}
\begin{align}
    I_{\text{I}} &= g_{\text{I}} G_{a}^{p_a}   G_{b}^{p_b} (V - E_{\text{I}}), \\
    I_{\text{FT}} &= g_{\text{FT}}  G_{c}^{p_c}   G_{d}^{p_d} (V - E_{\text{O}}), \\
    I_{\text{ST}} &= g_{\text{ST}}  G_{e}^{p_e}   G_{f}^{p_f} (V - E_{\text{O}}), \\
    I_{\text{SS}} &= g_{\text{SS}}  G_{g}^{p_g}  (V - E_{\text{O}}), \\
    I_{\text{L}} &= g_{\text{L}}  (V - E_{\text{L}}),
\end{align}
\label{eq:gate}
\end{subequations}
where $g_i$, $i\in\{\text{I},\text{FT}, \text{ST}, \text{SS}, \text{L}\}$, are the peak conductances, $E_j$, $j\in\{\text{I}, \text{O}, \text{L}\}$, are the reversal potentials of the currents, $G_k$, $k \in \{a,b,c,d,e,f,g \}$, are the gating variables and $p_k$ are their exponents. As suggested by \cite{Anderson_1989}, we assume that the three outward currents have the same reversal potential. The dynamics of a gating variable $G_k$ follow
\begin{equation}
    \frac{\text{d} G_k}{\text{d} t} = (G_{k\infty} - G_k)/ \tau_{G_k}.
    \label{eq:infinity}
\end{equation}
The voltage dependence of its steady-state value $G_{k\infty}$ is given by a logistic function with slope-factor $\rho_k$ and half-maximal voltage $V_{{1/2}_k}$,
\begin{equation}
    G_{k\infty}(V) = \frac{1}{1 + \exp((V_{{1/2}_k}-V)/\rho_k)},
    \label{eq:ginfinite}
\end{equation}
and the voltage dependence of its time constant $\tau_{G_k}$ is given by a Gaussian,
\begin{equation}
    \tau_{G_k}(V) = C_{\text{base}_k} + C_{\text{amp}_k} \exp\left(\frac{-(V_{\text{max}_k} - V)^2}{\sigma_k^2}\right).
    \label{eq:timeconstant}
\end{equation} \\ Here $C_{\text{base}_k}$ is the base value of $\tau_{G_k}$, $C_{\text{amp}_k}$ specifies its maximum at $V = V_{\text{max}_k}$  and $\sigma_k$ is the width of the Gaussian.

To fit the models for the transmembrane currents (Eq.\ \eqref{eq:gate}), we extract data points from the voltage clamp experiments of \cite{Anderson_1989}, Fig.~5 in Ch.~19, using WebPlotDigitizer \citep{Webplotdigitizer}. We simultaneously fit all 57 parameters using the L-BFGS algorithm \citep{Zhu_Byrd_Lu_Nocedal_1997} to minimize the least-squared error between model and data. We apply the basin hopping algorithm \citep{Olson_Hashmi_Molloy_Shehu_2012} to avoid getting caught in local minima. After obtaining the parameters for the transmembrane currents, we choose the membrane capacitance $C_m$ such that an action potential has similar features as reported in \cite{Anderson_1985}. Concretely, we set $C_m=1~\text{pF}$ to ensure that (i) the inflection point of an action potential is close to $0~\text{mV}$ and (ii) it takes about $2.5~\text{ms}$ for an EPSP to generate an action potential, with the synaptic parameters detailed in the next section. This fits well with the capacity of a deaxonized spherical soma of diameter $5-10~\text{\textmu m}$ \citep{Anderson_1985} and a specific capacitance of $1~\text{\textmu F/cm}^2$ \citep{Gentet_Stuart_Clements_2000}.

\subsection{Synapse Model}
\cite{Anderson_1985} found a voltage threshold of approximately $+20~\text{mV}$ for synaptic transmitter release in a scyphozoan synapse. In our network model we thus assume that when a neuron reaches this threshold from below (which happens during action potentials), excitatory postsynaptic currents (EPSCs) are evoked in the postsynaptic neurons, after a synaptic delay. The model EPSCs \citep{Gerstner_Kistler_Naud_Paninski_2014} rise with time constant $\tau_{\text{rise}}$, decay initially fast with time constant $\tau_{\text{fast}}$ and then tail off with a larger time constant $\tau_{\text{slow}}$,
\begin{equation}
        I_{\text{EPSC}} (t) =g_{\text{syn}} \left[1- e^{-t/\tau_{\text{rise}}}\right] \left[a  e^{-t/\tau_{\text{fast}}} + (1-a)  e^{-t/\tau_{\text{slow}}}\right]\Theta \left(t\right) \text{max}\left[(E_{syn}-V), 0 \right].
    \label{eq:synapse}
\end{equation}
Here, $E_{syn}$ is the current's reversal potential, $a$ the fraction of fast decay and $\Theta \left(t\right)$ the Heaviside theta function. The maximum function implements a synaptic rectification reported by \cite{Anderson_1985}: at potentials above the reversal potential synaptic currents do not reverse but stay zero.
The sum of individual EPSCs evoked in a postsynaptic neuron at times $t_0, t_1 \dots, t_n$ yields the total synaptic current $I_{\text{syn}}$ entering Eq. \eqref{eq:model},
\begin{equation}
    I_{\text{syn}}(t) = \sum\limits_{i=0}^n  I_{\text{EPSC}}\left(t- t_i \right) .
\end{equation}

\subsection{Motor Nerve Net}
\begin{figure}[htb!]
    \centering
    \begin{overpic}[width=\textwidth]{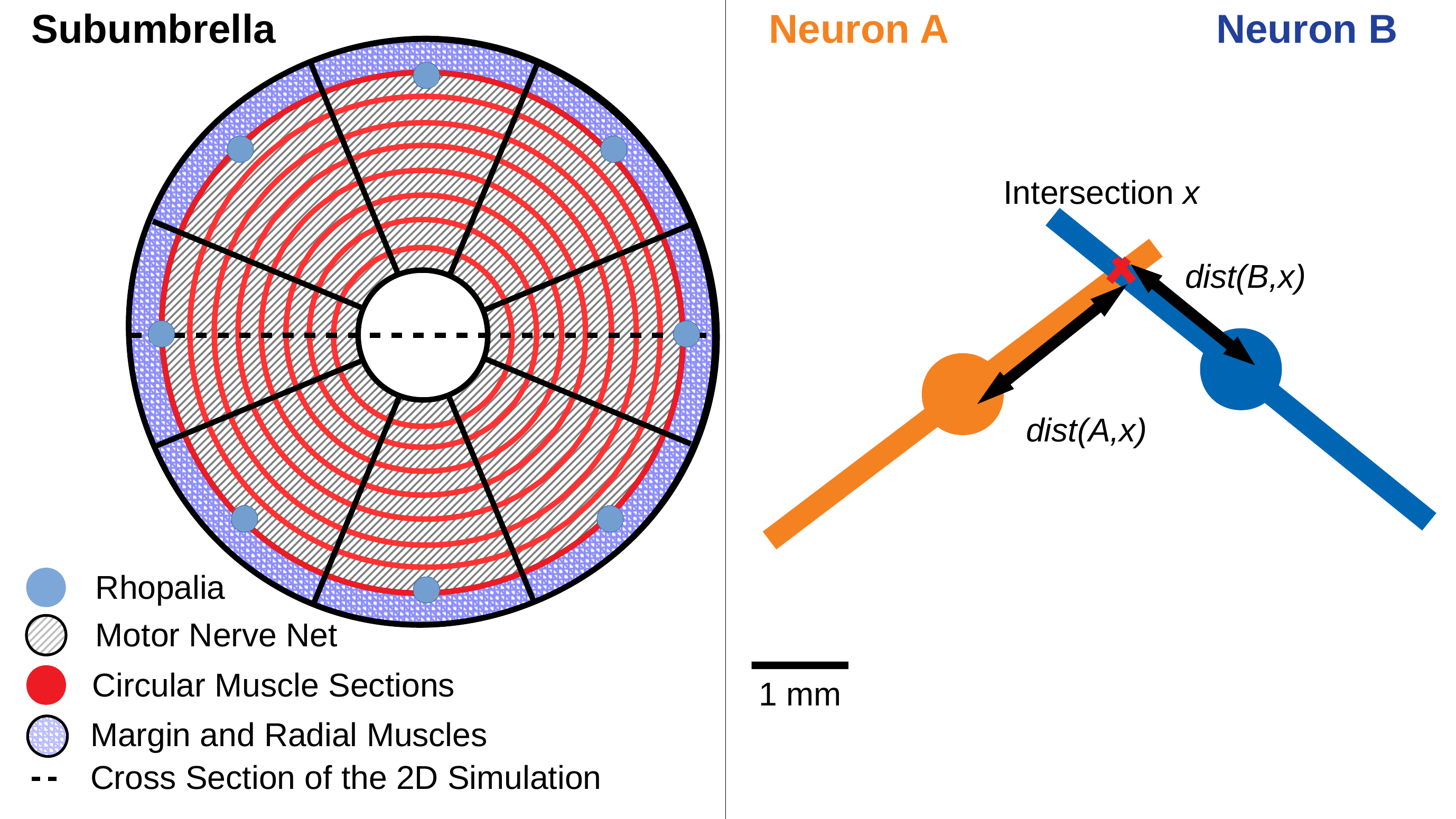}
        \put(45,2){A}
        \put(95,2){B}
    \end{overpic}

    \caption[Refractory]{\textbf{The jellyfish model.} 
    (\textit{A}) We model the jellyfish subumbrella as a disc with radius $2.25~\text{cm}$. The MNN somata are embedded in an annulus with an outer radius of $2~\text{cm}$ and an inner radius of $0.5~\text{cm}$ (gray hatched), leaving the margin and the manubrium region void. We assume that the circular swimming muscles (thick red) form discrete sections of concentric circles around the manubrium. The centers of these sections are aligned with the positions of the rhopalia. The DNN is distributed over the annulus between manubrium and margin and the margin with width $0.25~\text{cm}$ (blue hatched). For the hydrodynamics simulations, we use a cross section of the jellyfish as indicated by the dashed line. (\textit{B}) We model the spatial geometry of MNN neurites as line segments (rods) and assume that the soma is in their center (discs). Two neurons are synaptically connected if their neurites overlap. The transmission delay is a function of the distances between the somata and the intersection of their line segments (Eq.~\eqref{eq:dist}). 
    }
    
    \label{fig:Model}
\end{figure}

To capture the spatial properties of the nerve nets we model the spatial geometry of MNN neurons as line segments of length $5\text{mm}$ and assume that the soma is in their center (see Fig.~\ref{fig:Model} B). Two neurons are synaptically connected if their neurites overlap. The transmission delay between them is given by the constant synaptic delay of $0.5~\text{ms}$ and the distances between the somata and the intersection $x$ of the line segments (in cm). The total delay $\rho$ of two neurons with somata $A$ and $B$ is then given by
\begin{equation}
    \rho = 0.5 \si{\milli\second} +( \text{dist}(A,x) + \text{dist}(B,x) ) \; v,
    \label{eq:dist}
\end{equation}
where $v = 2\text{ms}/{\text{cm}}$.
This delay varies between $0.5$ and $1.5~\text{ms}$ and is constant for a given pair of neurons as observed by \cite{Anderson_1985}.

\begin{figure}[htb!]

    \begin{overpic}[width=0.95\textwidth]{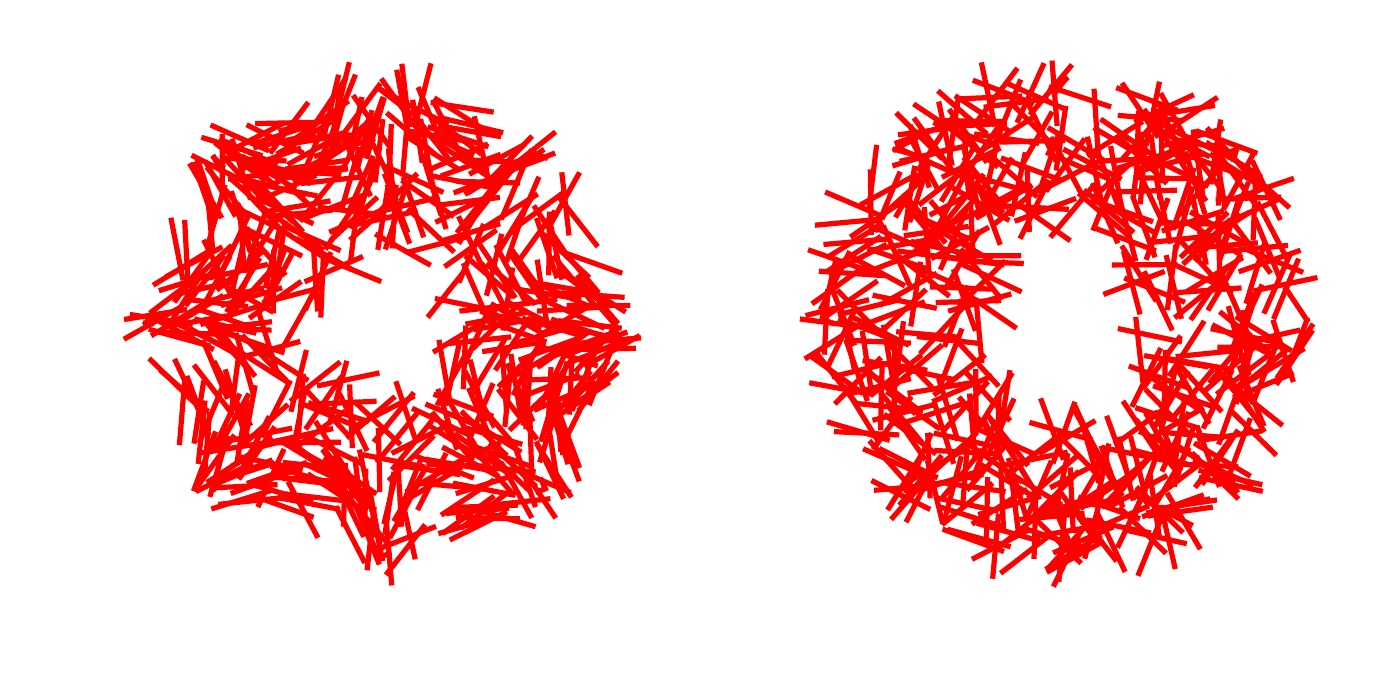}
        \put(6, 45){Von Mises}
        \put(52, 45){Uniform}
        \put(7,5){A}
        \put(53,5){B}
    \end{overpic}
    \caption[Networks]{\textbf{Example MNN models.} Two MNNs consisting of 500 neurons with von Mises (\textit{A}) or uniformly distributed (\textit{B}) neurite orientation.
    }
    \label{fig:Networks}
\end{figure}

We assume that neurons in the MNN are randomly placed on the subumbrellar surface. The orientation $\phi$ of their neurites relative to a straight line from the center of the bell to an (arbitrary) rhopalium is drawn from a von Mises distribution, with parameters depending on the position of the neuron,
\begin{equation}
    f(\phi | d, \alpha) = \frac{e^{8(d-0.5) \: \text{cos}(\phi - 3 \alpha)}}{2 \pi I_0(8(d-0.5))}.
    \label{eq:vonmise}
\end{equation}
Here, $d$ is the distance of the neuron from the center (in cm) and $\alpha$ is its polar angle relative to the line from the center to the rhopalium. $I_0(k) = \sum\limits_{m=0}^\infty \frac{1}{m!\Gamma(m+1)} (\frac{k}{2})^{2m}$ is the modified Bessel function of order zero, normalizing the expression. Eq.~\eqref{eq:vonmise} implements the  position dependence of the orientation distribution reported in \textit{Aurelia aurita} \citep{HORRIDGE_1954}, by (i) changing the variance of orientations with $d$ and (ii) changing the mean of the orientation distribution with $\alpha$. For comparison, we also consider networks with randomly uniform neurite orientation. Fig.~\ref{fig:Networks} displays example networks with the two different types of orientation distributions.

\begin{figure}[htb!]
    \centering
    \begin{overpic}[width=0.45\textwidth, height = 0.45\textwidth]{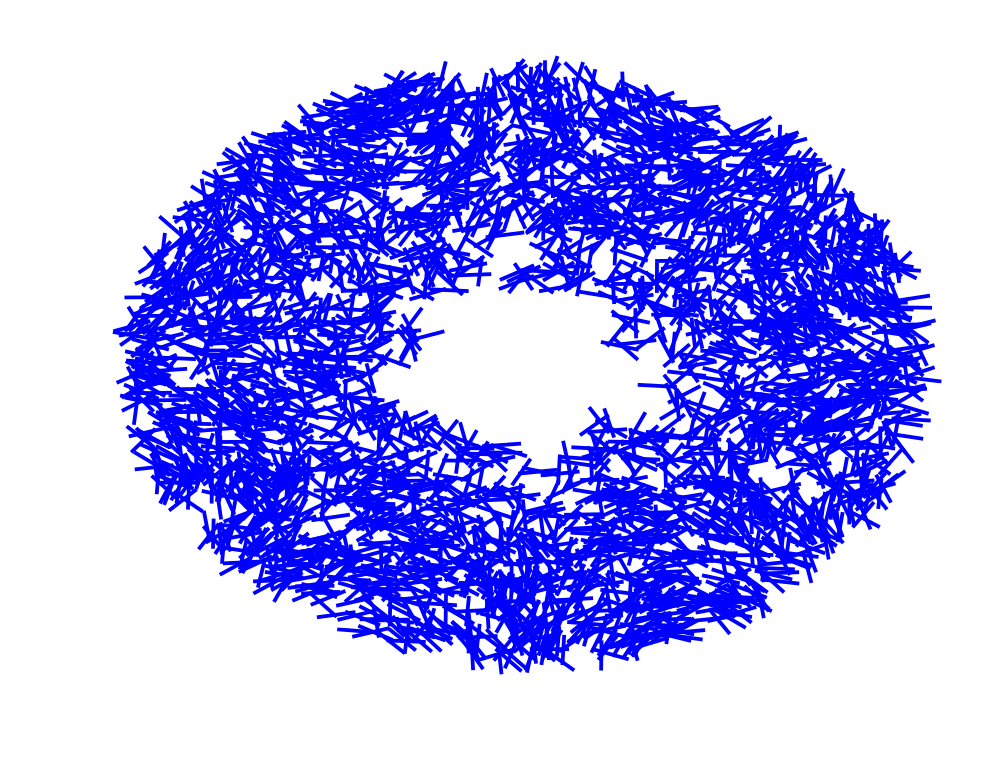}
        \put(0,5){A}
    \end{overpic}
    \begin{overpic}[width=0.45\textwidth, height=0.45\textwidth]{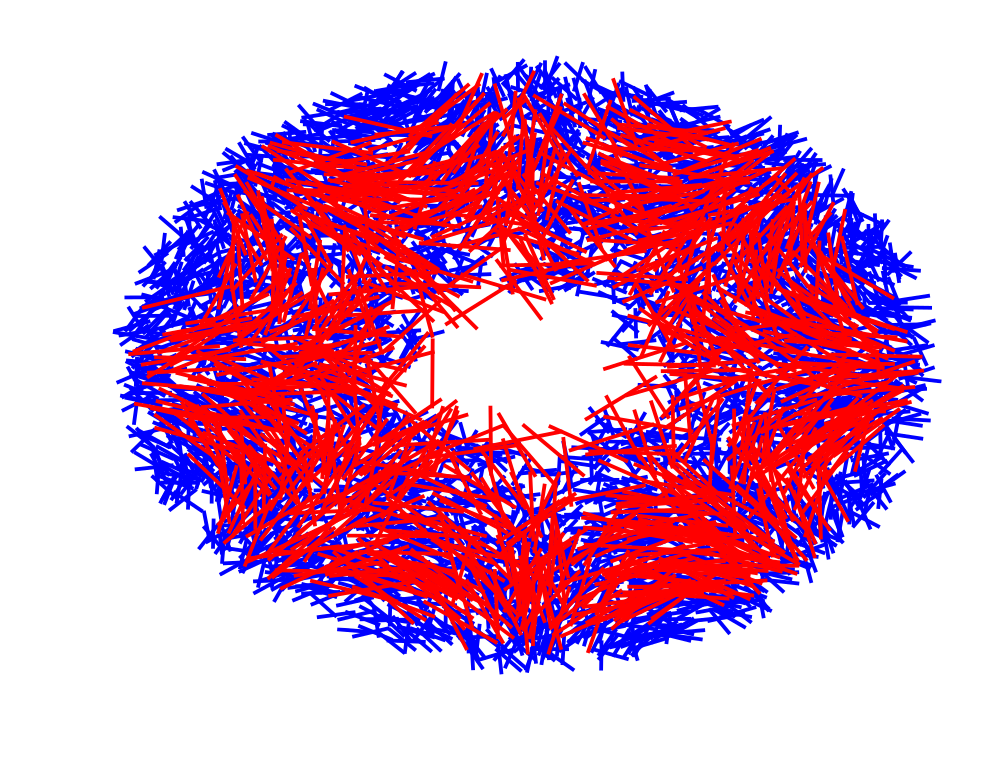}
        \put(0,5){B}
    \end{overpic}
    \caption[DNN]{\textbf{Example DNN model.} 
    (\textit{A}) A DNN with 3500 Neurons. (\textit{B}) The DNN (blue) and an MNN with 1000 neurons (red) displayed together. The DNN extends further into the bell margin.
    }
    
    \label{fig:DNN}
\end{figure}
\subsection{Diffuse Nerve Net}
We model the DNN similarly to the MNN, since little is known about it. In particular, we assume the same channel dynamics for DNN as for MNN neurons. There are, however, three main differences between the network models: First, the DNN extends into the bell margin \citep{Passano_Passano_1971}, which we take into account by increasing the maximum distance of the neurons from the center of the bell by $0.25~\text{cm}$ (blue hatched area in Fig. \ref{fig:Model}). Second, we set the overall length of DNN neurons to $2~\text{mm}$, in agreement with experimental observations \citep{Passano_Passano_1971}. Third, due to lack of data we assume random uniformly distributed neurite orientations. Fig.~\ref{fig:DNN} shows an example DNN network.

\subsection{Muscles}
\label{sec:muscles}
To model the activation of circular swimming muscles by MNN neurons (see Fig. \ref{fig:Model}), we follow a simple model for muscle force twitches used in \cite{Raikova_Aladjov_2002,Contessa_Luca_2012}: We assume that the time course of a muscle activation evoked by a single spike of an MNN neuron is given by
\begin{equation}
    a(t)= t^m e^{-kt}\Theta \left(t\right).
    \label{eq:twitch}
\end{equation}
We choose the rise and relaxation time parameters $m$ and $k$ such that the muscle activity duration is in the range of a variety of jellyfish species (see \cite{Satterlie_2015}, Tab.~2).

The force exerted by an activated muscle depends on its instantaneous extension. This effect prevents pathological muscle contraction by limiting the range of muscle activity. To incorporate the dependence, we adopt a simple model for force-length relationships \citep{Battista_Baird_Miller_2015}, assuming that the maximal force $F^j_I$ that a muscle fiber $j \in \{1, \dots, 64 \}$ of length $L^j_F$ can exert, is given by
\begin{equation}
    F^j_I(L^j_F) = F_O\exp{\left[-\left(\frac{L^j_F/L^j_O - 1}{\text{S}}\right)^2\right]}.
\end{equation}
Here, $L^j_O$ is the optimal length, $F_O$ the maximal force, which is generated at length $L^j_O$, and $\text{S}$ is a muscle-specific constant. $L^j_O$ is set to the length of the resting muscle. For simplicity, we do not include a force-velocity dependence in our model.

In summary, the force of a muscle fiber $j$ with length $L^j_t$ at time $t$ is in our model
\begin{subequations}
    \begin{align}
        f_j(t, L_t) = F^j_{I}(L^j_{t})\sum\limits_{i=0}^{n_j} a\left(t-t^j_i\right),  
    \end{align}
where $ t^j_0, t^j_1 \dots, t^j_{n_j} $ are the spike times of the MNN neurons innervating muscle $j$. We choose the constant $F_O$ such that
    \begin{align}
        \max\limits_{t,j} F_O \sum\limits_{i=0}^{n_j} a\left(t-t^j_i\right) = F_{\text{Norm}}
    \end{align}
    \label{eq:muscleforce}
\end{subequations}
after simulating the nerve net activity.
Hence, the muscle strength lies between 0 and $F_{\text{Norm}}$ after an excitation wave has passed through the MNN. All muscles are normalized in the same way, such that the relative strength between them stays constant independent of the number of neurons and the conduction speed.

The circular muscles of \textit{Aurelia aurita} are modeled as blocks of eight muscle units ordered radially in the area of each rhopalium. In total, we thus have 64 muscles (see Fig. \ref{fig:Model} A). We assume that a neuron is connected to one of those muscles if its  somatic position lies in the area covered by the muscle.

The radial muscles in the bell margin are modeled in the same manner as the circular ones. They are separated into eight blocks in the bell margin (see Fig.~\ref{fig:Model} A) and are innervated  by DNN neurons in the same way as the circular muscles are innervated by MNN neurons. Their activity is also governed by Eq.~\eqref{eq:twitch}- \eqref{eq:muscleforce} and they are also normalized in the same manner, independently of the circular muscles.

\subsection{Simulation of the Swimming Motion}
\subsubsection*{The Immersed Boundary Method}
To model the swimming behavior of the jellyfish we use the Immersed Boundary (IB) method \citep{Peskin_1972, Peskin_2002}. It was originally formulated to study flow patterns surrounding heart valves and has since been used for systems with intermediate Reynolds numbers,
\begin{equation}
    \text{Re} = \frac{\rho V L }{\mu},
    \label{eq:Re}
\end{equation}
of $10^{-1}$ to $10^3$. Here, $\rho$ and $\mu$ are the density and the viscosity of the surrounding fluid and $V$ and $L$ are the characteristic velocity and length of the problem \citep{Battista_Strickland_Miller_2017}.
In our simulations we set the maximal Reynolds number to approximately 250 by adjusting the viscosity of the fluid. This is in the range of Reynolds numbers calculated for swimming oblate Medusozoans \citep{Colin_Costello_2002} and yields a stable swimming motion in 2D simulations \citep{Herschlag_Miller_2011}. We use the IB2D package by \cite{Battista_Baird_Miller_2015, Battista_Strickland_Miller_2017, Battista_Strickland_Barrett_Miller_2017} to implement the simulation.

\subsubsection*{2D Jellyfish Geometry}
For our hydrodynamics simulations we develop a simple 2D construct, which is similarly shaped as 2D geometrical sections of \textit{Aurelia aurita} measured by \citep{Bajcar_Malacic_Malej_sirok_2008,McHenry_Jed_2003}. Our method of defining outlines allows in principle to create a wide variety of shapes including realistic cross sections of both prolate and oblate jellyfish while requiring only few parameters.
\begin{figure}[htb!]
    \centering
    \includegraphics[width=.8\textwidth]{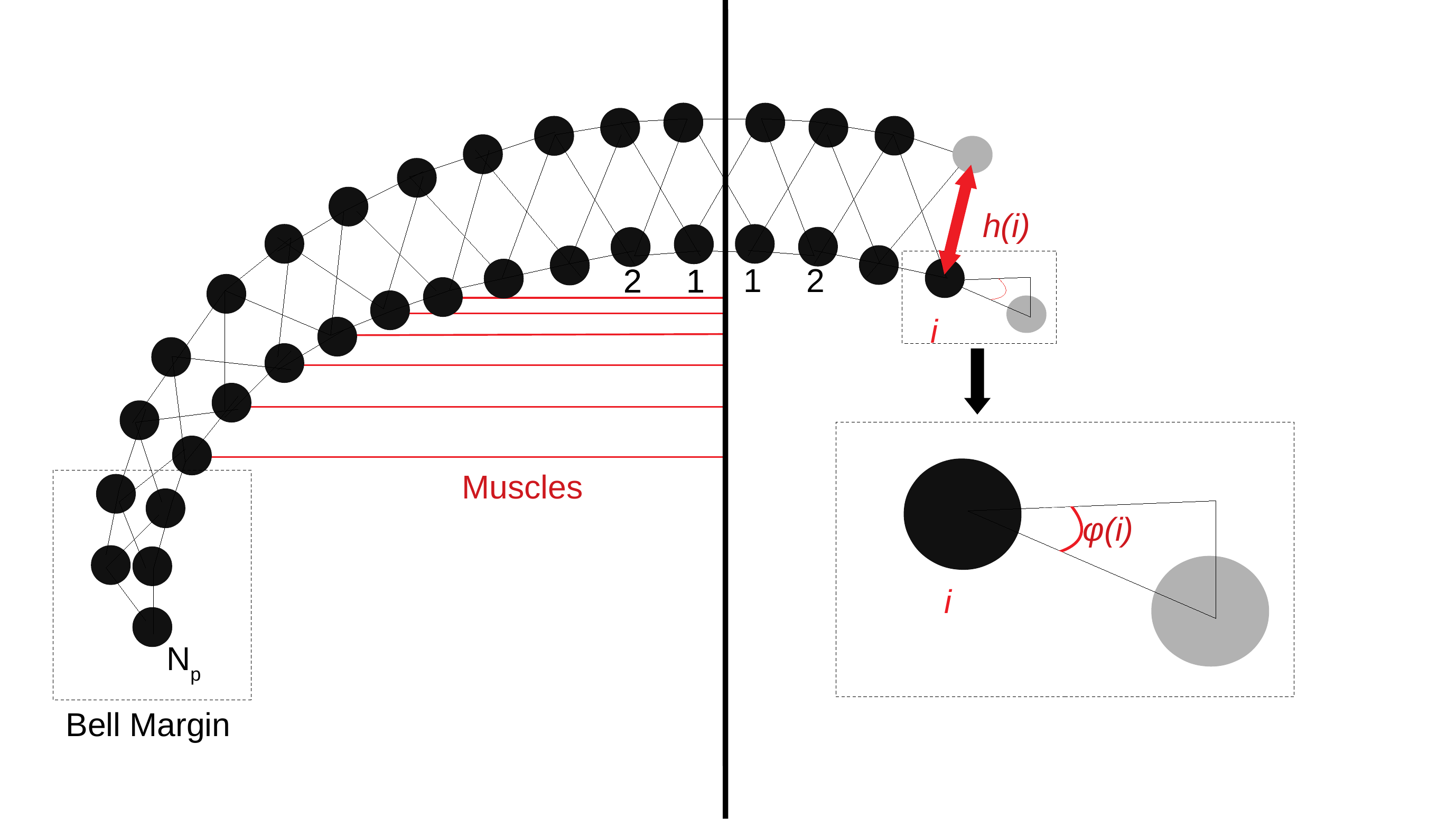}
    \caption[Refractory]{\textbf{The Jellyfish 2D sectional model.} The 2D structure consists of two rows of vertices, which are connected by damped springs (black lines). The placement of the vertices in the subumbrella (bottom row) depends only on the angle $\varphi(i)$ (Eq.~\eqref{eq:bell}). The vertices in the exumbrella (top row) are placed at a distance $h(i)$ (Eq.~\eqref{eq:height}) perpendicular to the curve traced by the bottom vertices. The circular muscles (red lines), which contract the bell, create a force (Eq.~\eqref{eq:muscleforce}) towards the imaginary center line of the jellyfish. No circular muscles are present at the center of the bell and the bell margin. 
    }
    \label{fig:Geometry}
\end{figure}

We define the relaxed shape of the subumbrella cross section with length $2 r$ by a series of $N_p$ vertices tracing a curve, on each half of the jellyfish. Specifically, the vertices are placed at constant distances $r/N_p$ from one another;  the negative angle $\varphi (i)$ between horizontal line and connection of $i$th and $i+1$th vertex (see Fig.~\ref{fig:Geometry}) decreases on the right hand side half with $i=0,...,N_p-1$ as
\begin{equation}
    \varphi (i) = -\alpha (1- p) \left(\frac{i}{N_p}\right)^{n_1} - \alpha p \left(\frac{i}{N_p}\right)^{n_2}.
    \label{eq:bell}
\end{equation}
Here, $\alpha$ (usually $\pi/2$) is the angle between the current orientation (center line) of the jellyfish and the horizontal line. The exponents $n_1$ and $n_2$ characterize the jellyfish's curvatures: the higher their values, the more oblate the jellyfish. $p$, a number between $0$ and $1$, characterizes the contribution of the two curvatures. To preserve the distance between the vertices, the first vertex is placed at half the usual distance (i.e. $r/(2 N_p)$) from the center of the subumbrella curve. Analogous expressions hold for the left hand side half. We note that for $n_1 = n_2 = 1$ the subumbrella is a semicircle with radius $2r/\pi$. 

The exumbrellar surface is defined by a series of vertices perpendicular to the subumbrella vertices (see Fig.~\ref{fig:Geometry}). Specifically, the $i$th exumbrellar vertex,  $i=1,...,N_p-1$, lies at a distance $h(i)$ to the $i$th subumbrellar vertex, perpendicular to the curve traced by the subumbrellar vertices. We model the height $h(i)$ of the jellyfish umbrella by base height plus a Gaussian hump
\begin{equation}
    h(i) = C_\text{base} (N_p -i)+ C_\text{amp} \exp{\left(  \frac{i^2}{\sigma^2}  \right)},
    \label{eq:height}
\end{equation}
where $C_\text{base}$ is the minimal height of the umbrella and $C_\text{amp}$ and $\sigma$ characterize the maximum height and the width of the umbrella's central hump.

\subsubsection*{2D Elastic Structure }\label{sec:jelly}
The jellyfish is an elastic structure filled with fluid; in particular the opening after a swimming contraction is a passive process \citep{Alexander_1964, Gladfelter_1972, Gladfelter_1973}. To incorporate this, we also construct the 2D cross section of the bell as an elastic structure filled with fluid \citep{Alexander_1964}: a set of damped springs run across the exumbrellar and the subumbrellar surfaces and connect the two surfaces defined by the vertices of the 2D cross section (see Fig.~\ref{fig:Geometry}). In the IB2D package the force on two vertices with coordinate vectors $\boldsymbol{X_1}, \boldsymbol{X_2}$ connected by a damped spring is defined by
\begin{equation}
    \boldsymbol{F_\text{s}} = k_s \left(1-  \frac{R_L}{||\boldsymbol{X_1} - \boldsymbol{X_2}||} \right) \left( \boldsymbol{X_1} - \boldsymbol{X_2} \right) + b_s \frac{\text{d}}{\text{d}t} ||\boldsymbol{X_1} - \boldsymbol{X_2}||,
    \label{eq:spring}
\end{equation}
where $R_L$ is the resting length, $k_S$ the spring stiffness and $b_S$ the damping coefficient.

Since the length of the 3D circular muscles and their radius are proportional we model them by muscles that are attached at subumbrellar vertices and exert the forces given by Eq.~\eqref{eq:muscleforce} directly towards the center line (see Fig.~\ref{fig:Geometry}, red).
To simulate the contraction of the radial muscles, we place DNN innervated muscles between neighboring vertices alongside the subumbrellar springs of the bell margin.

\section{Acknowledgments}
We thank Peter A.V. Anderson and Alexander P. Hoover for fruitful discussions and the German Federal Ministry of Education and Research (BMBF) for support via the Bernstein Network (Bernstein Award 2014, 01GQ1710).
\bibliography{ref}
\newpage
\section{Supplements}
\begin{table}[htbp!]
    \caption{Neuron model parameters}
     \centering
     \begin{tabular}[t]{|c|c|c|}
         \hline
         Variable & Value & Unit \\
        \hline
        $C_m$    & $1$& $\text{pF}$ \\ 
        $g_\text{I}$    & $345$&$\text{nS}$ \\
        $g_\text{FT}$    & $39.8$& $\text{nS}$ \\ 
        $g_\text{ST}$    & $27.2$& $\text{nS}$\\
        $g_\text{SS}$    & $10.8$& $\text{nS}$\\
        $g_\text{L}$    & $953$& $\text{pS}$\\
        $E_\text{I}$    & $76.7$&  mV\\
        $E_\text{O}$    & $- 84.6$ & mV\\
        $E_\text{L}$    & $-70$&  mV\\
        $p_a$    & $1.77$&  \\
        $p_b$    & $4.82$&  \\
        $p_c$    & $8.64$&  \\
        $p_d$    & $2.51$&  \\
        $p_e$    & $3.85$&  \\
        $p_f$    & $1.15$&  \\
        $p_g$    & $1$&  \\
        $V_{{1/2}_a }$    & $-2.02$& mV \\
        $V_{{1/2}_b }$    & $-10.94$& mV\\
        $V_{{1/2}_c }$    & $2.4$&  mV\\
        $V_{{1/2}_d }$    & $2.21\cdot 10^{-2}$&  mV\\
        $V_{{1/2}_e }$    & $10.65$&  mV\\
        $V_{{1/2}_f }$    & $-10.01$& mV\\
        $V_{{1/2}_g }$    & $48.58$&  mV\\
        $\rho_a $    & $3.99$& mV \\
        $\rho_b $    & $-13.03$& mV \\
        $\rho_c $    & $22.55$&  mV\\
        $\rho_d $    & $-8.97$&  mV\\
        $\rho_e $    & $26.43$&  mV\\
        $\rho_f $    & $-4.57$&  mV\\
        \hline
     \end{tabular}
     \begin{tabular}[t]{|c|c|c|}
        \hline
        Variable & Value& Unit \\
        \hline
       
        $\rho_g $    & $22.41$&  mV\\ 
        $C_{\text{base}_a} $    & $5.2 \cdot 10^{-1}$& ms \\ 
        $C_{\text{base}_b} $    & $1.3$&  ms\\ 
        $C_{\text{base}_c} $    & $1.65 \cdot 10^{-1}$& ms \\ 
        $C_{\text{base}_d} $    & $2.73$& ms \\ 
        $C_{\text{base}_e} $    & $1.13$& ms \\ 
        $C_{\text{base}_f} $    & $7.66$     & ms\\ 
        $C_{\text{base}_g} $    & $10.43$& ms \\ 
        $C_{\text{amp}_a} $    & $4.66\cdot 10^{-1}$ &ms \\
        $C_{\text{amp}_b} $    & $2.42\cdot 10^{-1}$ &ms \\
        $C_{\text{amp}_c} $    & $7.51$ &ms \\
        $C_{\text{amp}_d} $    & $10$&ms \\
        $C_{\text{amp}_e} $    & $16.64$ & ms\\
        $C_{\text{amp}_f} $    & $2$& ms \\
        $C_{\text{amp}_g} $    & $4.96$&  ms\\
        $V_{\text{max}_a} $    & $-5.87\cdot 10^{-1}$& mV \\
        $V_{\text{max}_b} $    & $2.68\cdot 10^{-1}$& mV \\
        $V_{\text{max}_c} $    & $-35.22$& mV \\
        $V_{\text{max}_d} $    & $-29.96$&  mV\\
        $V_{\text{max}_e} $    & $-12.71$& mV \\
        $V_{\text{max}_f} $    & $-34$& mV \\
        $V_{\text{max}_g} $    & $-39.93$& mV \\
        $\sigma_a $    & $1$& mV \\
        $\sigma_b $    & $6.62$&  mV\\
        $\sigma_c $    & $23.12$&  mV\\
        $\sigma_d $    & $15.13$&  mV\\
        $\sigma_e $    & $43.6$& mV \\
        $\sigma_f $    & $20$& mV\\
        $\sigma_g $    & $29.88$& mV\\ 
        \hline
     \end{tabular}
      
\end{table}
\begin{table}[htbp!]
    \caption{Synapse model parameters}
     \centering
     \begin{tabular}[t]{|c|c|c|}
         \hline
         Variable & Value & Unit \\
        \hline
        $g_\text{syn}$    & $75$& $\text{nS}$\\ 
        $\tau_\text{rise}$    & $20$& ms\\
        $\tau_\text{fast}$    & $3$& ms\\ 
        $\tau_\text{slow}$    & $6$& ms\\
        $a$    & $9.57\cdot 10^{-1}$& \\
        $E_\text{syn}$    & $4.32$& mV\\ 

        \hline
     \end{tabular}

\end{table}

\begin{table}[htbp!]
    \caption{Muscle model parameters}
     \centering
     \begin{tabular}[t]{|c|c|c|}
         \hline
         Variable & Value & Unit \\
        \hline
        $m$    & $1.075$ &\\ 
        $k$    & $2.15\cdot 10^{-2}$& \\

        $\text{S}$    & $0.4$& \\
        $F_\text{Norm}$ for circular muscles    & $0.4$& N  \\
        $F_\text{Norm}$ for radial muscles & $0.8$& N \\

        \hline
     \end{tabular}

\end{table}

\begin{table}[htbp!]
    \caption{Geometry parameters}
     \centering
     \begin{tabular}[t]{|c|c|c|}
         \hline
         Variable & Value & Unit \\
        \hline
        $\alpha$    & $\pi/2$& \\ 
        $p$    & $0.5$& \\
        $N_p$    & $ 224$& \\
        $n_1$    & $1$& \\
        $n_2$    & $2$& \\
        $C_\text{base}$    & $0.5$& mm\\
        $C_\text{amp}$    & $6$& mm\\
        $\sigma$    & $3000$& \\
        $k^\text{surface}_s$    & $2 \cdot 10^{7}$& N/m \\
        $k^\text{internal}_s$    & $8 \cdot 10^{7}$& N/m \\ 
        $b_s$    & $2.5$&  kg/s\\

        \hline
     \end{tabular}

\end{table}

\begin{table}[htbp!]
    \caption{Fluid Simulation parameters}
     \centering
     \begin{tabular}[t]{|c|c|c|}
         \hline
         Variable & Value & Unit \\
        \hline
        $\mu$    & $0.005$& $\text{Ns}/\text{m}^2$ \\
        $\rho$    & $1000$& $\text{kg}/\text{m}^2$\\
        Time step    & $10^{-5}$& s\\
        x-length of Eulerian grid & $0.06$& m\\
        y-length of Eulerian grid & $0.08$& m\\
        x-grid size & $180$& \\
        y-grid size & $240$& \\

        \hline
     \end{tabular}

\end{table}
\clearpage

\end{document}